\documentclass[fleqn,usenatbib]{mnras}

\usepackage{newtxtext,newtxmath}
\usepackage[T1]{fontenc}
\usepackage{ae,aecompl}
\usepackage{lscape}

\usepackage{graphicx}	% Including figure files
\usepackage{amsmath}	% Advanced maths commands
\title[$AKARI$ NEPW Field: Photo-z]{Photometric Redshifts in the North Ecliptic Pole Wide Field based on a Deep Optical Survey with Hyper Suprime-Cam}

\author[Simon C.-C. Ho et al.]{
Simon C.-C. Ho$^{1}$\thanks{E-mail: simonsimon259@gapp.nthu.edu.tw},
Tomotsugu Goto$^{1}$,
Nagisa Oi$^{2}$,
Seong Jin Kim$^{1}$,
\newauthor
Matthew A. Malkan$^{3}$,
Agnieszka Pollo$^{4,5}$,
Tetsuya Hashimoto$^{1,6}$,
Yoshiki Toba$^{7,8,9}$,
\newauthor
Helen K. Kim$^{3}$,
Ho Seong Hwang$^{10}$,
Hyunjin Shim$^{11}$,
Ting-Chi Huang$^{12,13}$,
\newauthor
Eunbin Kim$^{10}$,
Ting-Wen Wang$^{1}$,
Daryl Joe D. Santos$^{1}$,
and Hideo Matsuhara$^{13}$
\\
$^{1}$Institute of Astronomy, National Tsing Hua University, Hsinchu city, Taiwan\\
$^{2}$Faculty of Science Division II, Tokyo University of Science, Tokyo, Japan\\
$^{3}$Department of Physics and Astronomy, UCLA, 475 Portola Plaza, Los Angeles, CA 90095-1547, USA\\
$^{4}$National Centre for Nuclear Research, ul. Pasteura 7, 02-093 Warsaw, Poland\\
$^{5}$Astronomical Observatory of the Jagiellonian University, ul. Orla 171, 30-244 Cracow, Poland \\
$^{6}$Centre for Informatics and Computation in Astronomy (CICA), \\ National Tsing Hua University, 101, Section 2. Kuang-Fu Road, Hsinchu, 30013, Taiwan (R.O.C.)\\
$^{7}$Department of Astronomy, Kyoto University, Kitashirakawa-Oiwake-cho,
Sakyo-ku, Kyoto 606-8502, Japan\\
$^{8}$Academia Sinica Institute of Astronomy and Astrophysics, 11F of
Astronomy-Mathematics Building, AS/NTU, No.1, Section 4,\\ Roosevelt Road,
Taipei 10617, Taiwan\\
$^{9}$Research Center for Space and Cosmic Evolution, Ehime University, 2-5
Bunkyo-cho, Matsuyama, Ehime 790-8577, Japan\\
$^{10}$Korea Astronomy and Space Science Institute, 776 Daedeokdae-ro, Yuseong-gu, Daejeon 34055, Republic of Korea\\
$^{11}$Department of Earth Science Education, Kyungpook National University, 80 Daehak-ro, Buk-gu, Daegu 41566, Republic of Korea\\
$^{12}$Department of Space and Astronautical Science, Graduate University for Advanced Studies, SOKENDAI, Shonankokusaimura, \\
Hayama, Miura District, Kanagawa 240-0193, Japan\\
$^{13}$Institute of Space and Astronautical Science, Japan Aerospace Exploration Agency, 3-1-1 Yoshinodai, Chuo-ku, Sagamihara,\\
Kanagawa 252-5210, Japan\\}
\pubyear{2015}
\begin{document}
\label{firstpage}
\pagerange{\pageref{firstpage}--\pageref{lastpage}}
\maketitle
\begin{abstract}
The $AKARI$ space infrared telescope has performed near- to mid-infrared (MIR) observations on the North Ecliptic Pole Wide (NEPW) field (5.4 deg$^2$) for about one year. $AKARI$ took advantage of its continuous nine photometric bands, compared with NASA\textquotesingle s $Spitzer$ and WISE space telescopes, which had only four filters with a wide gap in the MIR.
The $AKARI$ NEPW field lacked deep and homogeneous optical data, limiting the use of nearly half of the IR sources for extra-galactic studies owing to the absence of photometric redshifts (photo-zs). 
To remedy this, we have recently obtained deep optical imaging over the NEPW field with 5 bands ($g$, $r$, $i$, $z$, and $Y$) of the Hyper Suprime-Camera (HSC) on the Subaru 8m telescope.  We optically identify AKARI-IR sources along with supplementary $Spitzer$ and WISE data as well as pre-existing optical data. In this work, we derive new photo-zs using a $\chi^2$ template-fitting method code ($Le$ $Phare$) and reliable photometry from 26 selected filters including HSC, $AKARI$, CFHT, Maidanak, KPNO, $Spitzer$ and WISE data. We take 2026 spectroscopic redshifts (spec-z) from all available spectroscopic surveys over the NEPW to calibrate and assess the accuracy of the photo-zs. At z < 1.5, we achieve a weighted photo-z dispersion of {$\sigma_{\Delta{z/(1+z)}}$} = 0.053 with $\eta$ = 11.3\% catastrophic errors.
\end{abstract}

%%%%%%%%%%%%%%%%% BODY OF PAPER %%%%%%%%%%%%%%%%%%
\section{Introduction}
Redshift is a crucial observable in studies of galaxies and in cosmology \citep[ e.g.,][]{Ilbert2006, Tanaka2018, Salvato2019}. Spectroscopic measurements are the best way to determine the redshifts of extragalactic sources. However, in practice, it is hard to carry out spectroscopic observations for a huge number of targets because of the enormous time and expense required. While many current and future surveys provide photometric data, the spectra of faint galaxies are not easy to obtain. Alternatively, calculating photometric redshift (photo-z) is an efficient and powerful way to get redshift information for a very large number of galaxies. Accordingly, the availability of large photometric datasets has been making photo-z estimates an essential component in modern extragalactic astronomy and cosmology studies.

To understand the cosmic star-formation history, it is important to investigate infrared (IR) emission because more than half of the UV-optical energy generated by star formation activity is obscured by dust and re-emitted in IR  \citep[e.g.,][]{Goto2010, Goto2017}.  This is a reason why the $AKARI$ space infrared telescope was launched in 2006. It immediately started a legacy survey--the NEP survey program-- covering 5.4 deg$^2$ using its unique 9 filter bands \citep{Matsuhara2006}. The NEP-Wide (NEPW) survey detected and catalogued 114,000 near- to mid-IR (NIR to MIR) sources published by \citet{Kim2012}.
However, 20-30\% of the NIR ($N2$, $N3$ and $N4$ band) sources have remained optically unidentified, because the follow-up optical imaging surveys were not deep enough,  \citep[e.g., the depth at $r$-band of the Canada-France-Hawaii Telescope CFHT was $\sim$ 25.6 mag at rest frame;][]{Hwang2007,Jeon2010,Kim2012}. The redshift and IR luminosity of these sources have never been measured. Galaxies at  z < 2 generally require both optical and NIR photometry for reliable photo-z derivation. Therefore, deep imaging with 5 bands ($g$, $r$, $i$, $z$, and $Y$) over the entire NEPW field was obtained using the Subaru/Hyper Suprime-Cam \citep[HSC;][Oi et al. 2020, in press]{Miyazaki2012, Miyazaki2018, Goto2017}. After data reduction with the recent pipeline, a  multi-band photometric catalogue was constructed (Kim et al. 2020, in press). To perform photo-z estimation, here we use the panchromatic data in this catalogue, from the optical $u^{*}$-band (CFHT)  to the W2 band of the  Wide-field Infrared Survey Explorer \citep[WISE;][]{Wright2010}. %and Photodetecting Array Camera and Spectrometer \citep[PACS;][]{Poglitsch2010} data where available (Kim et al, in prep.).

To compute photo-z in this work, spectral energy distribution (SED) fitting was performed by using the PHotometric Analysis for Redshift Estimate \citep[$Le$ $Phare$;][]{Arnouts1999, Ilbert2006} code. We present photo-z and associated probability distribution for all the sources in the NEPW catalogue. We also discuss the physical properties of faint IR sources in the NEPW field.  

This paper is composed as follows; we describe our data sets in Section \ref{sec:data}. Photo-z calculation with the multi-band data set using SED fitting is presented in Section \ref{sec:SED}. 
%Photo-z calculation with the multi-band data sets using Machine Learning is discussed in Section \ref{sec:MLphotoz}. 
Discussion of source properties is provided in Section \ref{sec:discuss}, followed by the conclusions in Section \ref{sec:conclusion}. Throughout this paper, we adopt the AB magnitude system, and assume a cosmology with H$_0$ = 70 kms$^{-1}$Mpc$^{-1}$, {$\Omega_\Lambda$} = 0.7, and\ {$\Omega_\text{M}$} = 0.3 \citep{Spergel2003}.

\section{THE DATASETS}\label{sec:data}

In this section, we present a brief description of the data we used in this work: summary of the source matching process, the multi-band photometry and the spectroscopic redshift (spec-z) data in the NEPW field. The multiwavelength data were combined for the IR sources from the $AKARI$'s NEPW survey \citep{Matsuhara2006, Lee2009, Kim2012} cross-matched with the deep Subaru/HSC optical data (Kim et al. 2020, in press).

\subsection{Optical Identification of the AKARI Sources}\label{opticaliden}

This work is based on the deep optical imaging survey carried out by the Subaru/HSC to cover the entire NEPW field of $AKARI$ \citep[][Oi et al. 2020, in press]{Goto2017}. The matching process was done by Kim et al. 2020 (in press). For a detailed matching process, please refer to Kim et al. 2020 (in press).

\citet{Kim2012} presented a photometric catalogue of IR sources based on the NEPW survey of $AKARI$ by using its 9 photometric bands of the IRC from 2-25 $\mu$m.
In the NIR bands, the $N2$ filter reaches a depth of $\sim$20.9 mag, and the $N3$ and $N4$ bands reach 21.1 mag.
AKARI's NEPW survey covered a circular area of 5.4 deg$^{2}$ centred at the NEP, %(R.A.=18${h}$00${m}$00${s}$, Dec.=+66$^{\circ}$33${'}$38${''}$) 
using 9 continuous IRC filters ($N2$, $N3$, $N4$, $S7$, $S9W$, $S11$, $L15$, $L18W$, and $L24$). 
%corresponding to 2.4, 3.2, 4.1, 7, 9, 11, 15, 18, and 24 $\mu$m of the reference wavelength. 
The official pipeline version 6.5.3 was used to reduce the HSC data (Oi et al. 2020, in press). 
All the NEPW IR sources (135,157) were cross-matched against the Subaru/HSC optical data (using a matching radius determined by 3$\sigma$ of the mean positional offset), which are eventually divided into two categories: clean (91,861) and flagged (19,674) sources according to the Subaru/HSC flag information (Oi et al. 2020, in press; Kim et al. 2020, in press). Those flagged sources are with base\char`_ PixelFlags\char`_ flag\char`_ edge (Source is outside usable exposure region), base\char`_ PixelFlags\char`_ flag\char`_ bad (Bad pixel in the Source footprint), and base\char`_ PixelFlags\char`_ flag\char`_ saturatedCenter (saturated pixel in the source centre) equal to 1. At the same time, 23,622 sources remained unmatched to Subaru/HSC data (no-HSC sources). We excluded the flagged and no-HSC sources in this work because the photo-z estimation requires good optical photometry. Those no-HSC sources in the NEPW will be discussed by \citet{toba2020}.
%AKARI$\textquotesingle s 9-band photometry (Kim et al. in prep). 
Kim et al. 2020 (submitted) also used additional supplementary optical data previously obtained with SNUCAM \citep[$B$, $R$, and $I$;][]{Im2010} at the Maidanak Observatory \citep{Jeon2010} and with CFHT/MegaCam \citep[$u^{*}$, $g$, $r$, $i$, $z$;][]{Hwang2007, Oi2014}, even though they were not deep enough to identify all the IR sources. Nonetheless, they are still valuable for providing detailed optical photometric data points between HSC filters. The $u^{*}$-band data from CFHT/MegaCam covers the wavelength range ($\sim$0.3 $\mu$m) beyond the HSC filters. They also included the additional CFHT/MegaPrime $u$-band data (PI: Goto, T.) reduced/calibrated by \citet{Huang2020}, which filled out the incomplete areal coverage of the pre-existing $u^{*}$-band  from  the  CFHT/MegaCam.
  
\subsection{Spectroscopic Data}\label{sec:specdata}
Among the 91,861 clean sources, we have 2,026 sources with spec-z from various surveys ($AKARI$-HSC-specz sources). The spectroscopic data are from several observations with various telescopes/instruments in the optical, Keck/DEIMOS \citep[Goto et al. in prep. ;][]{Shogaki2018, Kim2018}, MMT/Hectospec \citep[HEC;][]{Shim2013}, WIYN/Hydra \citep[HYD;][]{Shim2013}, GTC \citep[Miyaji et al. in prep.;][]{Diaz-Tello2017, Krumpe2015}, in the NIR, Subaru/FMOS \citep{Oi2017}, and the NIR to mid-infrared (MIR) \citep[`SPICY'][]{Ohyama2008,Ohyama2018}.

Among 2,026 spec-z data, the largest number ($\sim$ 60\%) comes from two spectroscopic campaigns (HEC \& HYD) from \citet{Shim2013}. 
They targeted $AKARI$ IR sources based on the MIR fluxes at 11 $\mu$m ($S11<18.5$ mag) and at 15 $\mu$m ($L15<17.9$ mag). Additional $R$-band magnitude cuts (16 $<R<$ 21 for HYD, 16 $<R<$ 22.5 for HEC) were introduced to secure reasonable signal-to-noise (S/N) ratio based on the optical data from the Maidanak/SNUCAM and CFHT/MegaCam. Most of these flux-limited sources are various types of IR luminous galaxies.
The second largest group of spectra consisting of 20\% of the whole spectroscopic sample, was observed with Keck/DEIMOS over different years \citep[][Goto et al. in prep.]{Shogaki2018, Kim2018}. Most of these sources are IR-selected star-forming galaxies (SFGs).
Additional spectroscopic data obtained mostly on the NEP Deep (NEPD) area including published  \citep[e.g., `SPICY' galaxies,][]{Ohyama2008, Ohyama2018} and unpublished spectra (e.g. Miyaji et al. in prep.) were included.

\begin{figure}
\centering
	\includegraphics[width=\columnwidth]{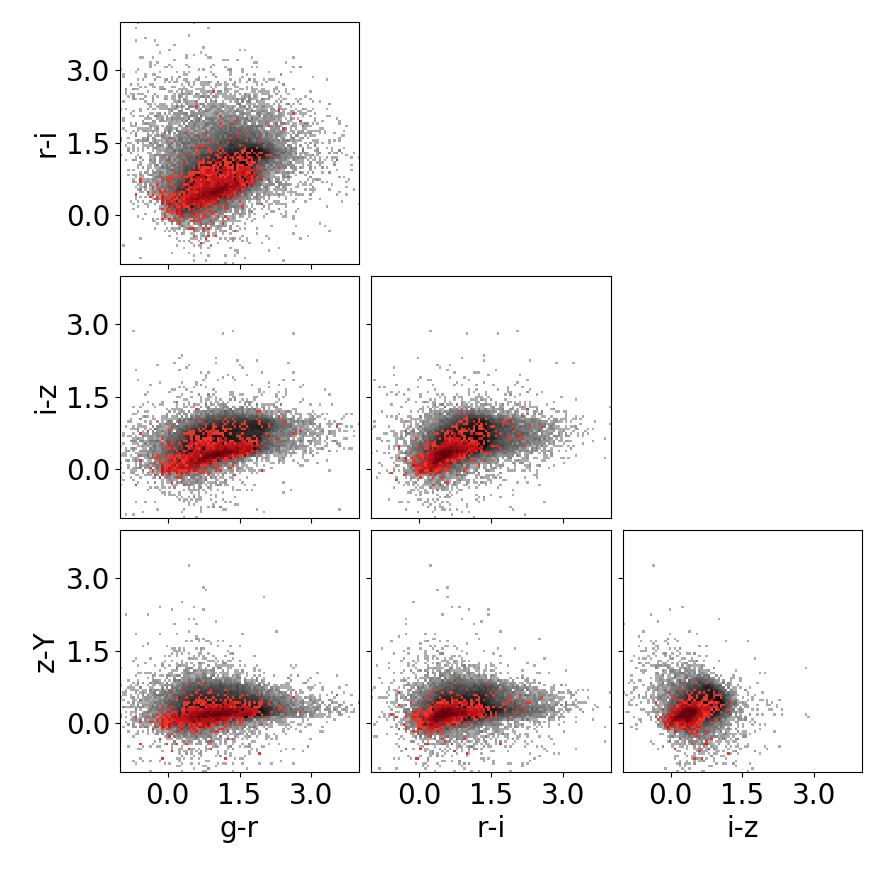}

    \caption{Colour-colour diagrams for the Subaru/HSC bands of all the galaxies in the catalogue. Red and black dots represent data with and without spec-z, respectively.}
    \label{fig:colorcolor}
\end{figure}

Fig. \ref{fig:colorcolor} shows the colour-colour diagrams for the Subaru/HSC bands of all the galaxies in the catalogue. Red and black dots represent objects with and without spec-z, respectively. Although the spectroscopic catalogue we used is a collection of spec-z from several spectroscopic campaigns, they show consistent colour distributions.  This is partly because they are all IR-selected galaxies in the NEPW region.

\begin{figure*}
\centering
	\includegraphics[width=150mm]{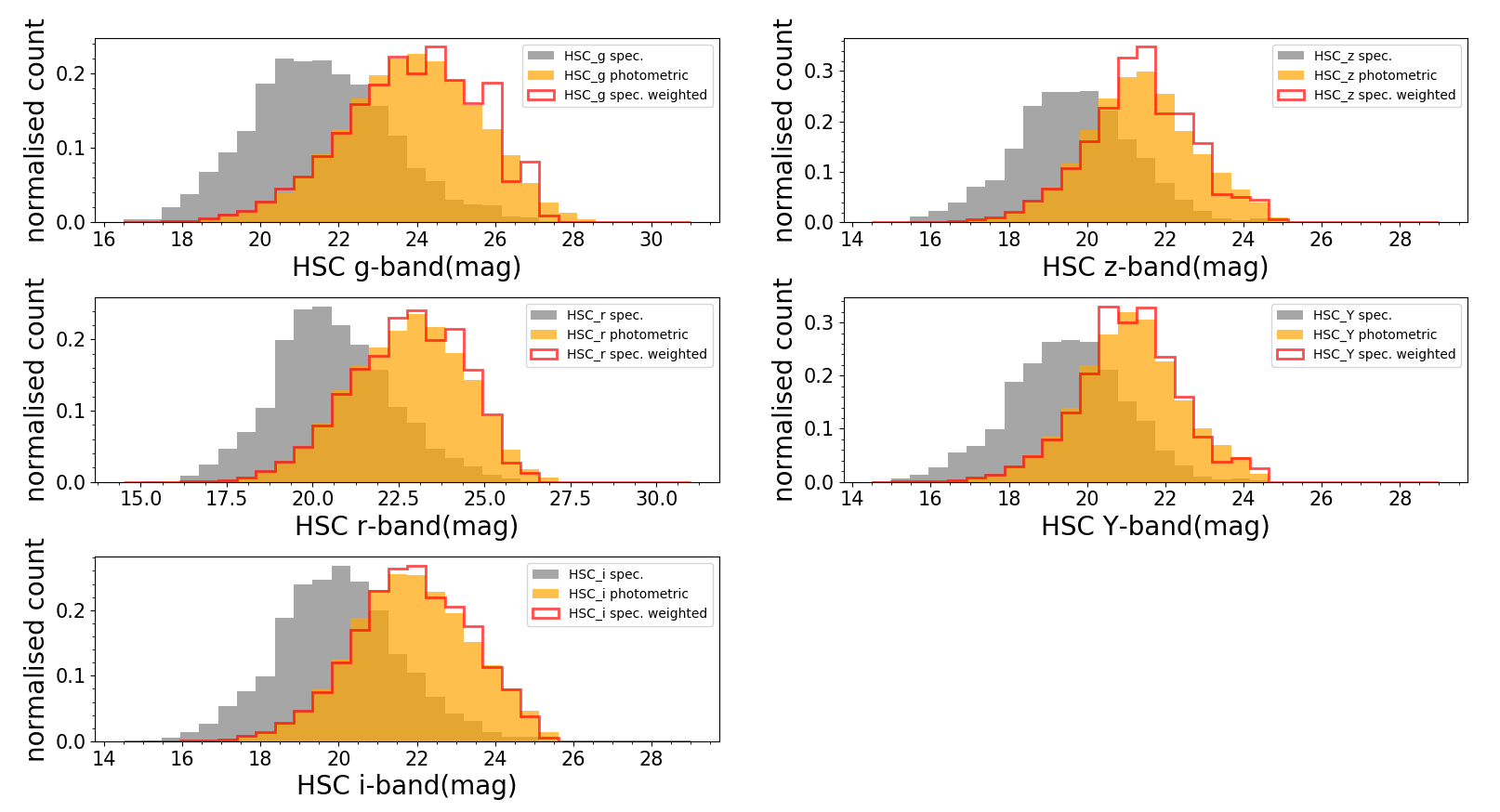}
    
    \caption{Magnitude distributions of Subaru/HSC $g$, $r$, $i$, $z$, $Y$-band galaxies. The grey filled histogram shows the magnitude distribution of the spec-z sample before weighting. The orange filled histogram is the magnitude distribution of the photometric sample. The red histogram is the magnitude distribution of the spec-z sample after weighting. Note that the histograms are normalised so that the area equals to 1.}
    \label{fig:magdistri}
\end{figure*}

Apart from source target bias, it is important to make sure there is no strong mismatch in magnitude between the spec-z catalogue and photometric catalogues. It is better to have a spec-z sample that extends to faint enough magnitudes so that the catalogue is useful for photo-z calibrations. To check this, we plot the histograms of the Subaru/HSC $g$, $r$, $i$, $z$, and $Y$-band galaxies' magnitude distribution (Fig. \ref{fig:magdistri}). The depth of our spectroscopic catalogue extends to $\sim$27 mag, $\sim$26 mag, $\sim$25 mag, $\sim$24 mag, $\sim$23 mag in Subaru/HSC $g$, $r$, $i$, $z$, $Y$-bands, respectively. However, the depth of the photometric catalogue is still deeper by $\sim$ 1 mag in every HSC band. We needed to check the magnitude distribution of spec-z sample so that it can reproduce that of all the photometric samples.

We adopted a nearest neighbour (NN) approach \citep{Lima2008} to reconstruct the magnitude distributions of the spec-z catalogue by computing weights for all spec-z sources. The NN algorithm is a non-parametric method used for classification and regression. As the algorithm does not take non-detection as an input, we can only consider the bands with the largest amounts of detections to prevent limiting the sample size. In our case, we use the Subaru/HSC $g$, $r$, $i$, $z$, $Y$-bands magnitudes of 2026 spec-z objects with 89,035 photometric sources as input. The algorithm then matches the magnitude distribution of 2026 spec-z objects with 89,035 photometric sources. 2026 weights for the spec-z sources are given as outputs. Weights are the values that indicate the contribution of a particular set of data. In this case, it is the contribution to magnitude distribution. We show the weighted magnitude distributions of the spec-z sample in Fig. \ref{fig:magdistri} along with the magnitude distributions of the photometric sample and the spec-z sample before weighting. In the figure, the weighted reconstructions of the magnitude distributions show fairly good agreement with the photometric sample. We present photo-z performance for both the weighted and unweighted cases. The weighted photo-z performance is shown in square brackets throughout this paper.

\subsection{Ancillary Near- to Mid-IR Data}
The $J$- and $H$-band catalogue \citep{Jeon2014} of the NEP region is obtained by using the Florida Multi-object Imaging Near-IR Grism Observational Spectrometer \citep[FLAMINGOS; ][]{Elston2006} on the Kitt Peak National Observatory (KPNO). On the other hand, the Wide-field Infrared Survey Explorer (WISE) mission surveyed the whole sky in four bands; W1 (3.4 $\mu$m), W2 (4.6 $\mu$m), W3 (12 $\mu$m), and W4 (22 $\mu$m), with spatial resolutions of 6.1, 6.8, 7.4 and 12.0 arcsec, respectively. We adopted the north \& south pole survey catalogue data release, which covered the NEP region over an area of $\sim$1.5 deg$^2$ \citep{Jarrett2011}. Moreover, the CFHT/ Wide-field InfraRed Camera (WIRCam) data is obtained from \citet{Oi2014}. The $Spitzer$ Infrared Array Camera (IRAC) provided an IRAC1 and IRAC2 two-band catalogue of the NEP field \citet{Nayyeri2018}. The observations covered 7.04 deg$^2$. 

\begin{figure*}
\centering
	\includegraphics[width=\columnwidth]{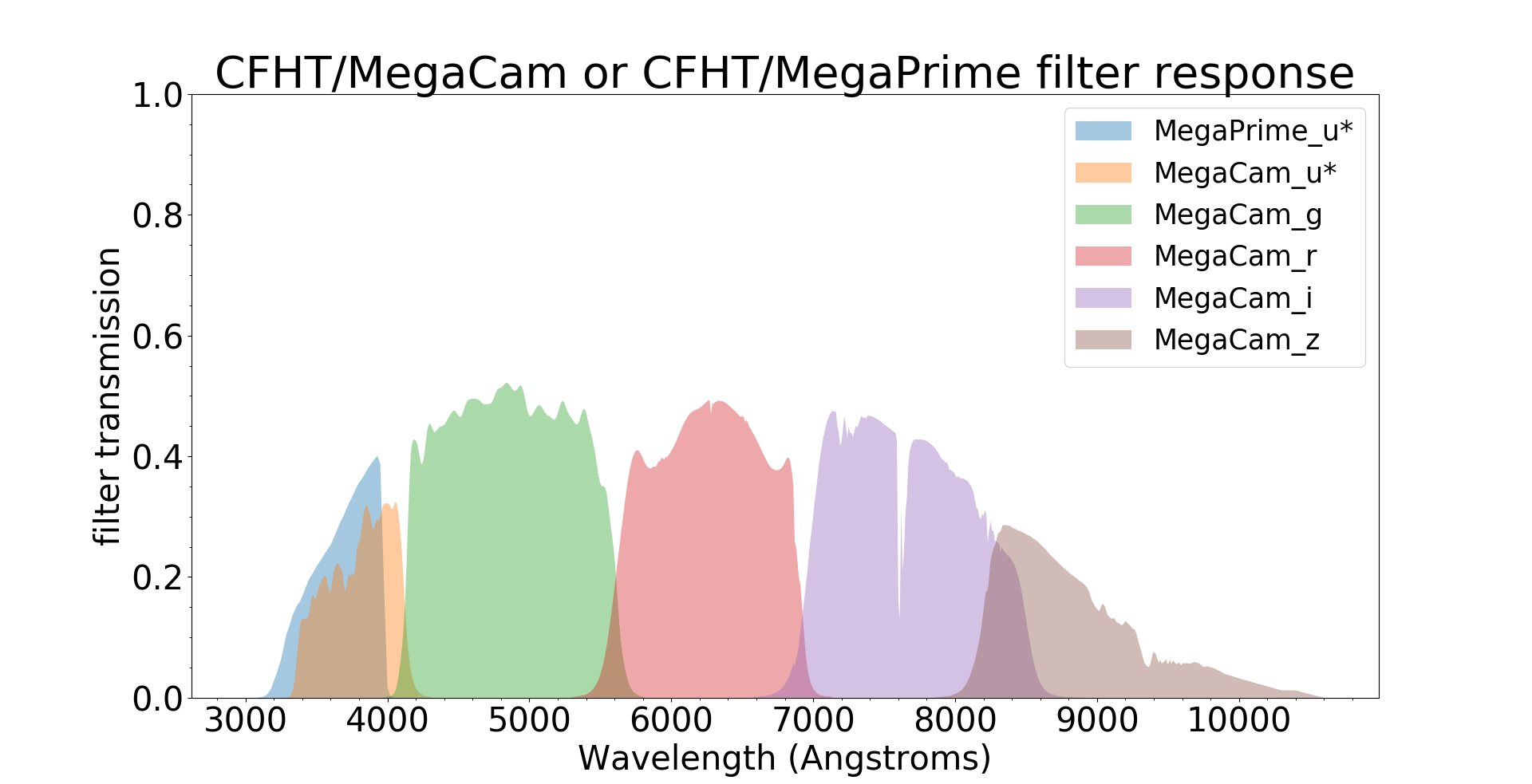}\includegraphics[width=\columnwidth]{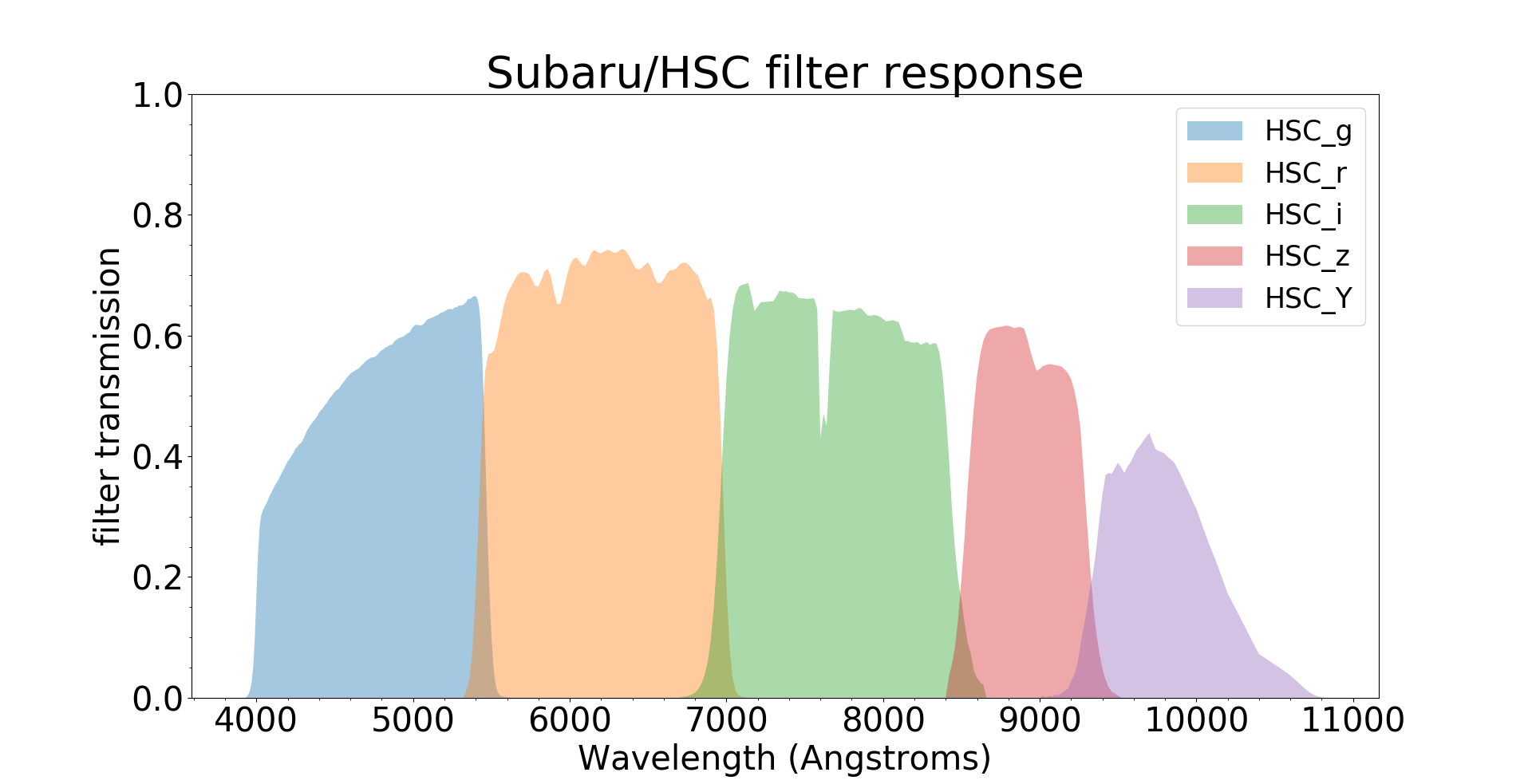}
	\includegraphics[width=\columnwidth]{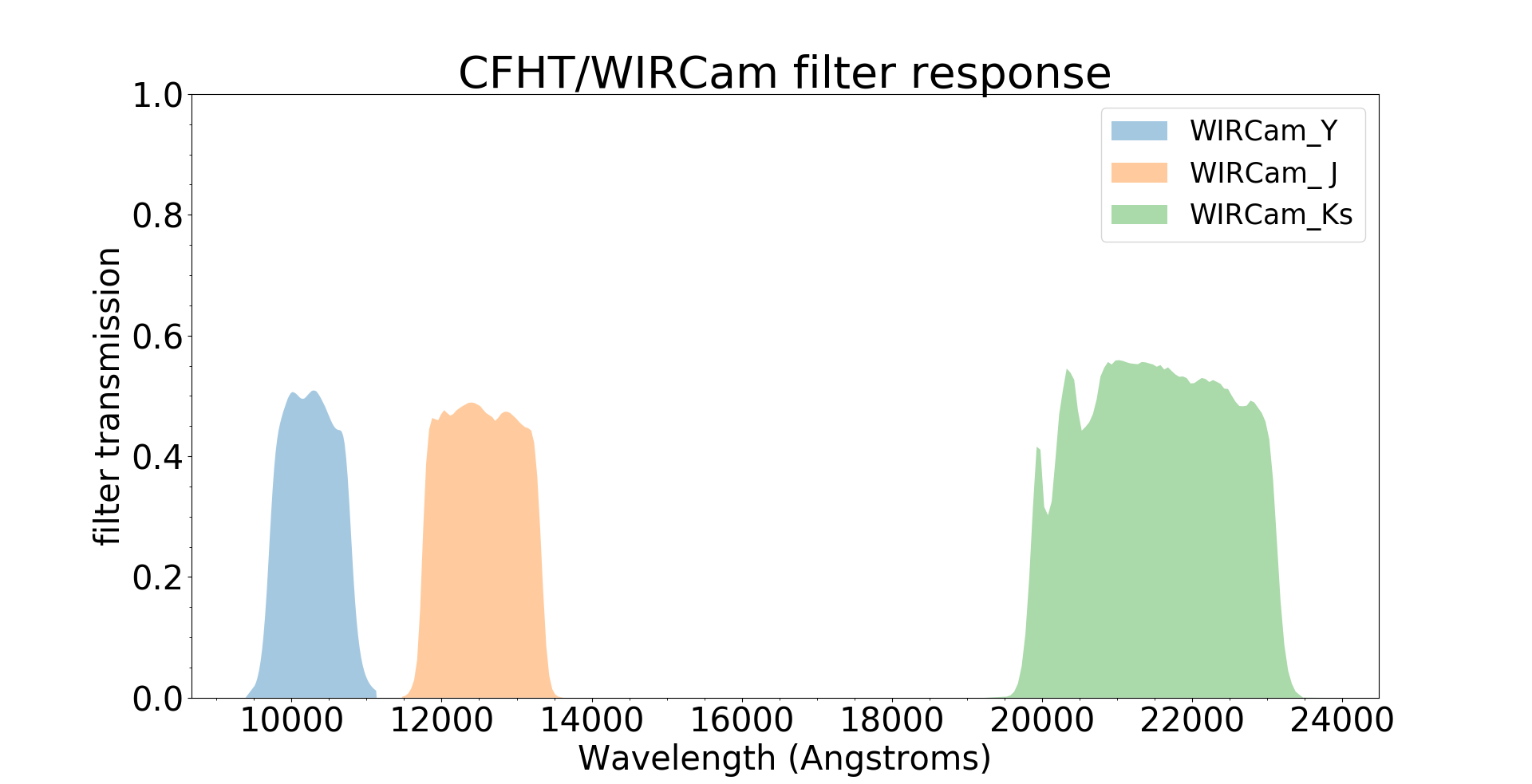}\includegraphics[width=\columnwidth]{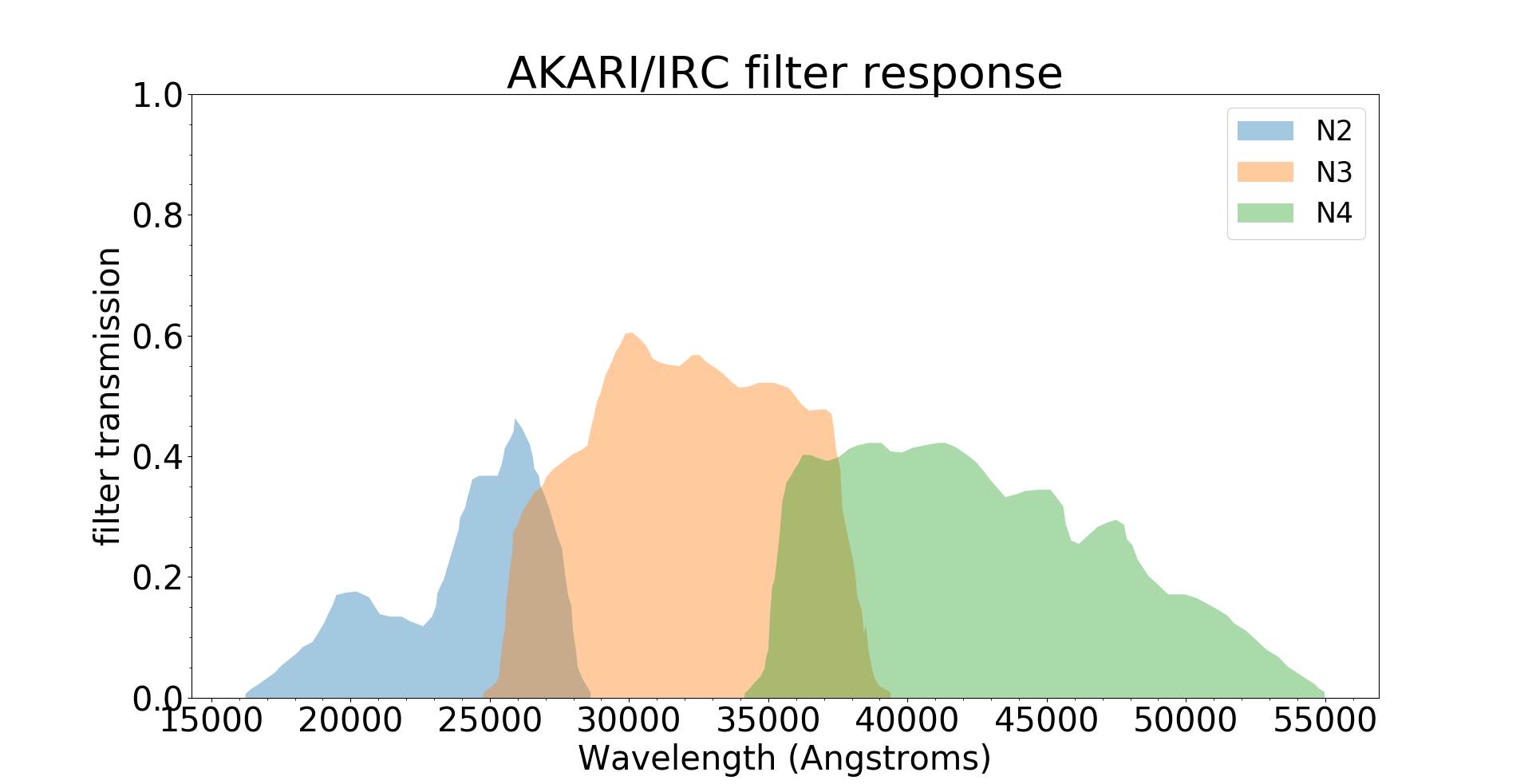}
	\includegraphics[width=\columnwidth]{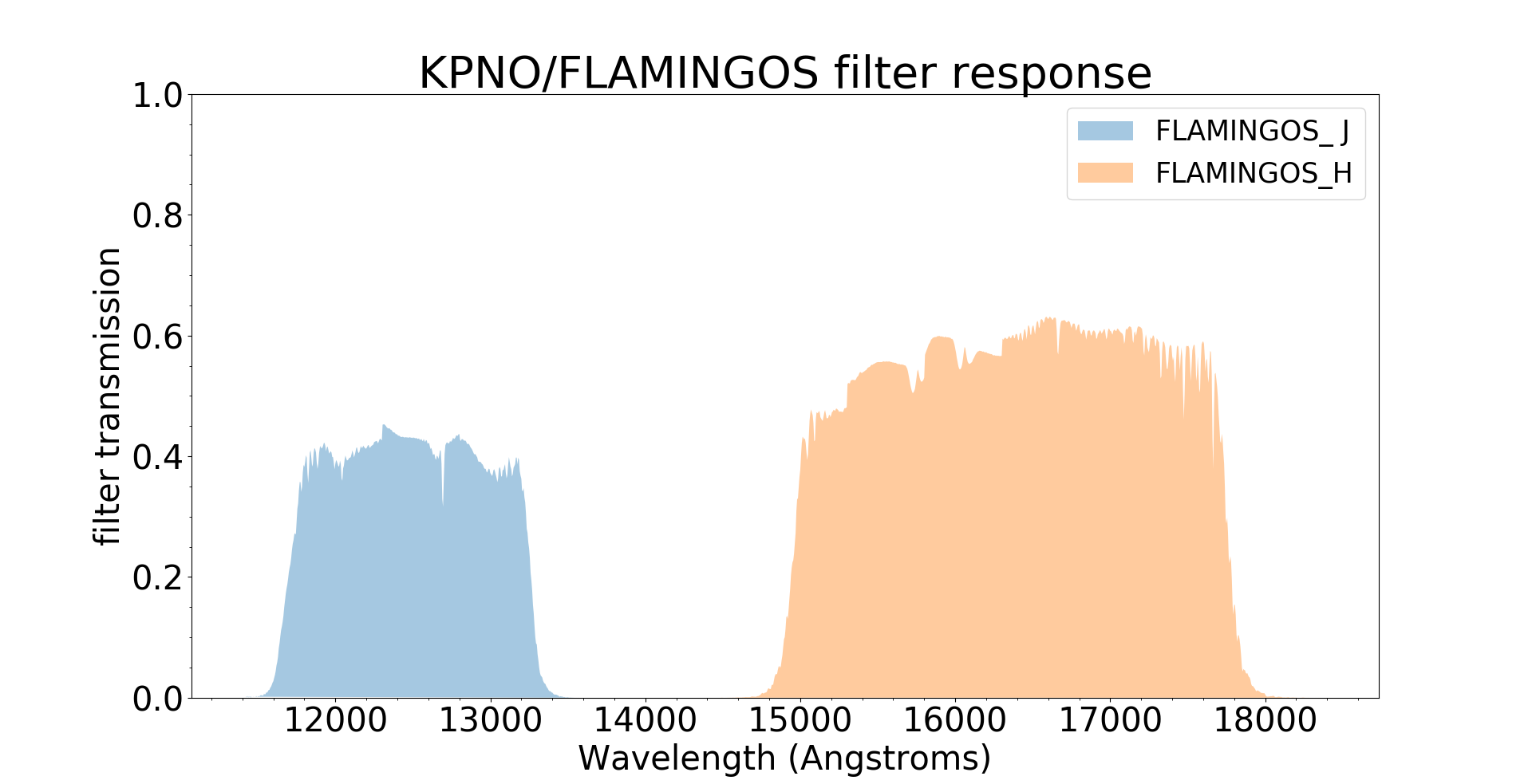}\includegraphics[width=\columnwidth]{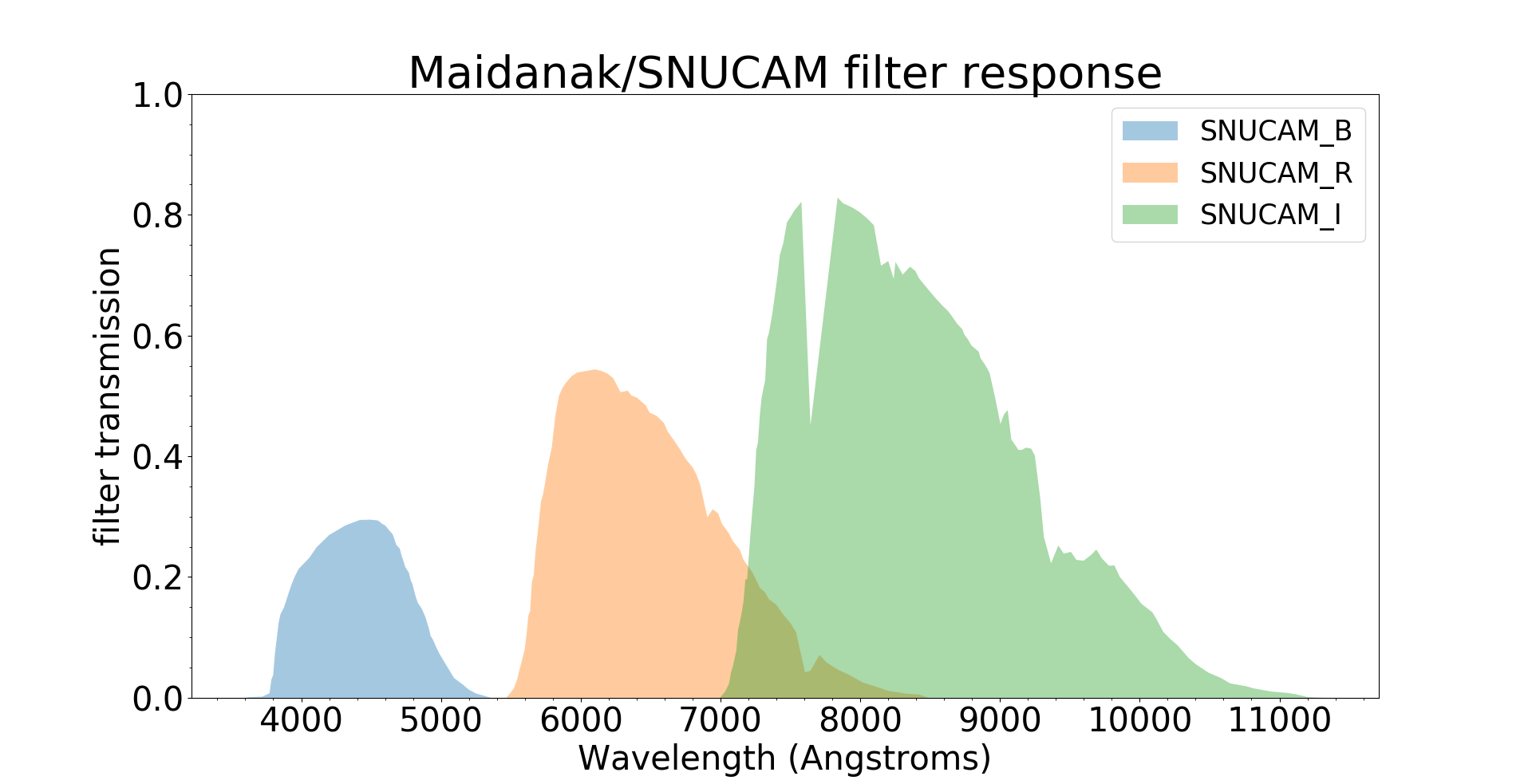}
	\includegraphics[width=\columnwidth]{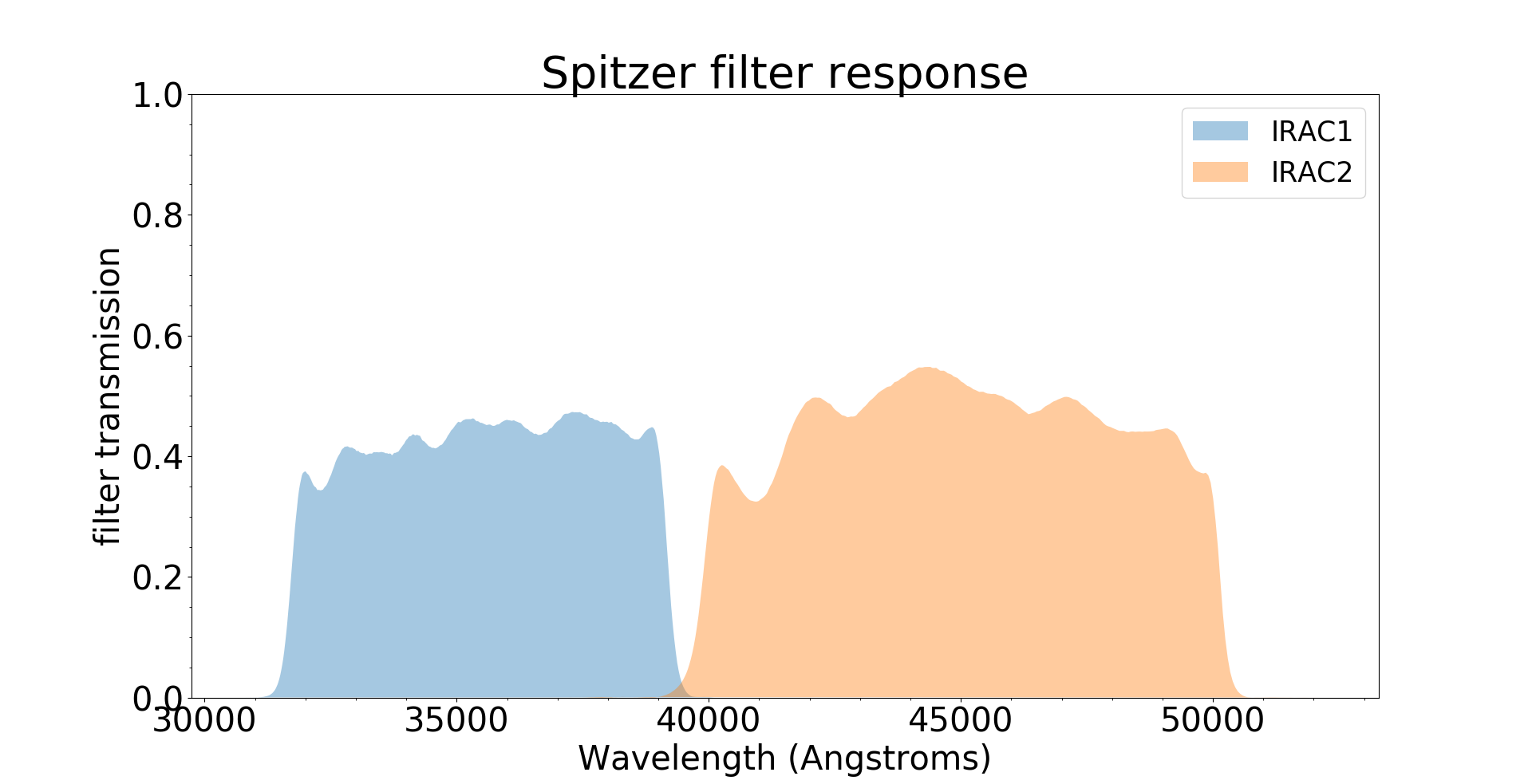}\includegraphics[width=\columnwidth]{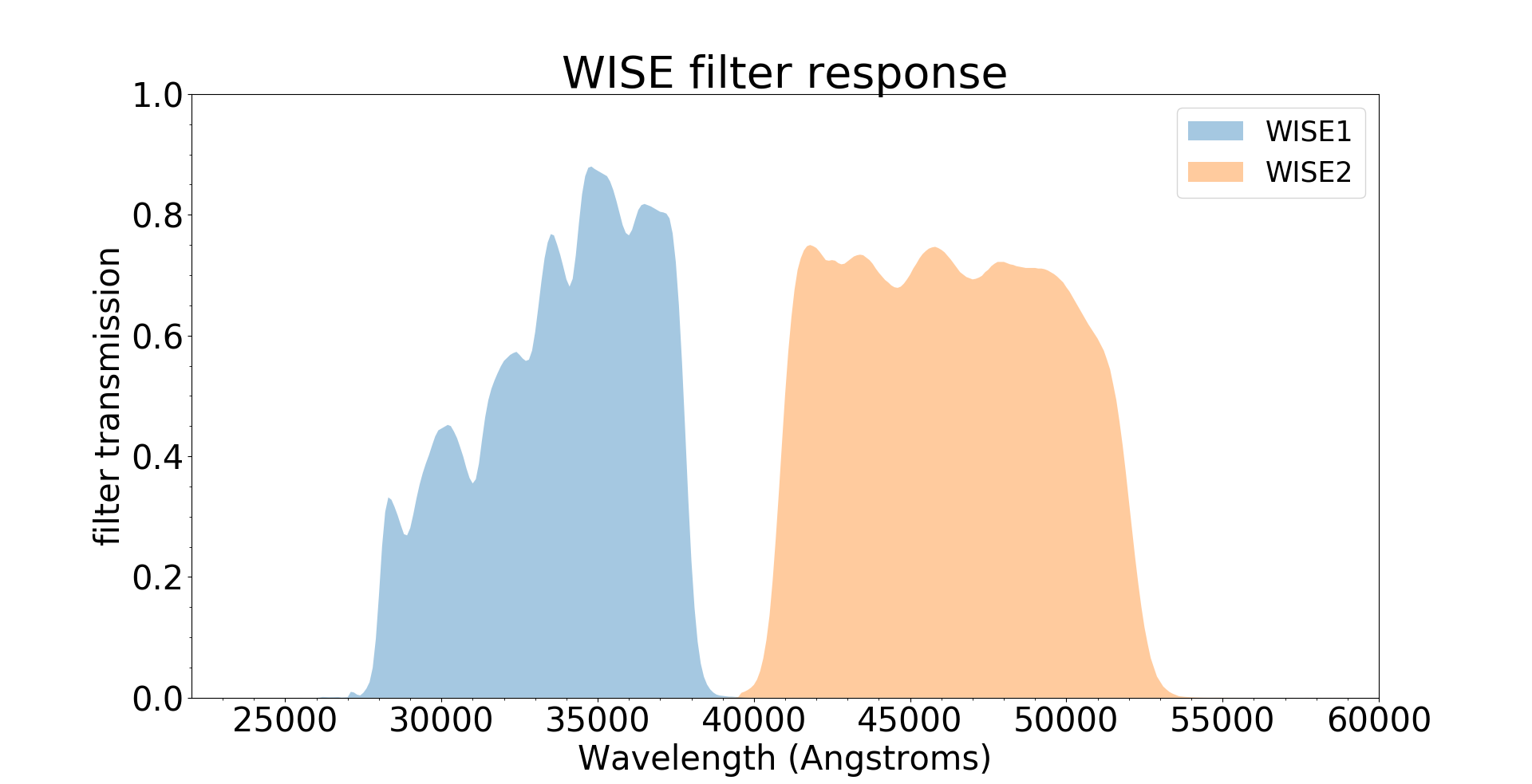}

    \caption{All the filters and the transmission curves we used in this work. The atmosphere (an airmass), telescope instruments, QE of the CCDs were taken into account together with the filter shapes to present these final filter transmissions.}
    \label{fig:filteresponse}
\end{figure*}

We integrated all the filter information to take airmass, telescope optics, Quantum Efficiency (QE) of the CCD into account, along with the filter responses. This information is summarised in Table \ref{tab:filterinfo}. We also include the depth and full width half maximum (FWHM) of the seeing size in this table. The descriptions of how objects are detected and measured in each band is summarised in APPENDIX \ref{detection}. The methodology to calculate the depths of each band is summarised in the APPENDIX \ref{depth}. The filter responses are also presented in Fig.~\ref{fig:filteresponse}.

\subsection{Data Selection}\label{sec:datacal}
In our catalogue, we found that in some cases, Subaru/HSC magnitudes are fainter than CFHT/MegaCam data by > 0.5 mag. This photometric discrepancy was found for about $\sim$ 1 \% of the sources. This discrepancy did not only originate from slightly different filter responses between the two bands but also from the deblending or multiple detections of structure of the source. In these cases, the Subaru/HSC band's flux of the bright sources is actually divided into several pieces due to the saturation/deblending. Deblending is caused by the source detection algorithm (of the Subaru/HSC pipeline) to separate an integrated signal, originated from many different places, eventually generated several split detections. As a result, the source can be observed as separate sources in one of the bands but only one source in other bands. Thus, there will be a false detection of the target and the photometry of these sources is incorrect.

\begin{figure*}
\centering
	\includegraphics[width=130mm]{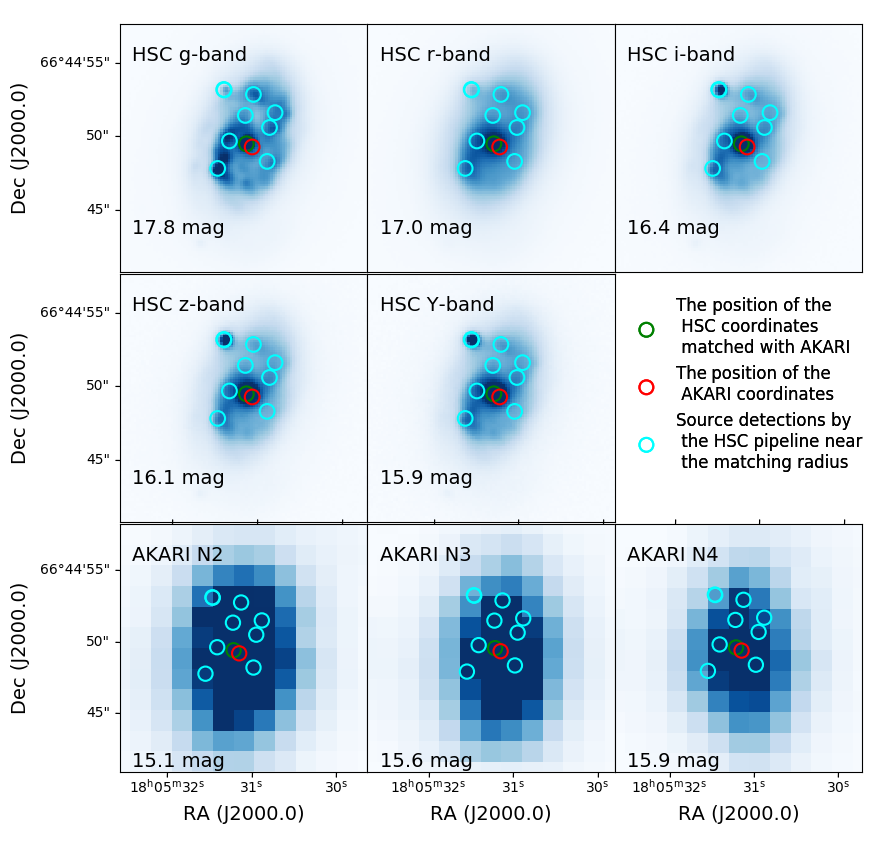}

    \caption{%This figure shows 
    A source detected in Subaru/HSC $g$, $r$, $i$, $z$, $Y$-bands and $AKARI$ $N2$, $N3$, $N4$-bands. The red circle shows the position of $AKARI$ detection (AKARI\char`_ ID:134105). The green circle shows the Subaru/HSC detection coordinates which matches the same $AKARI$ source. The cyan circles show the positions of Subaru/HSC detections (by the Subaru/HSC pipeline). The radius of the circles is 1 arcsec. Note that we apply arbitrary scaling to the images at each panel to make the structure clearer.}
    \label{fig:83827.pdf}
\end{figure*}

\begin{figure}
\centering
	\includegraphics[width=\columnwidth]{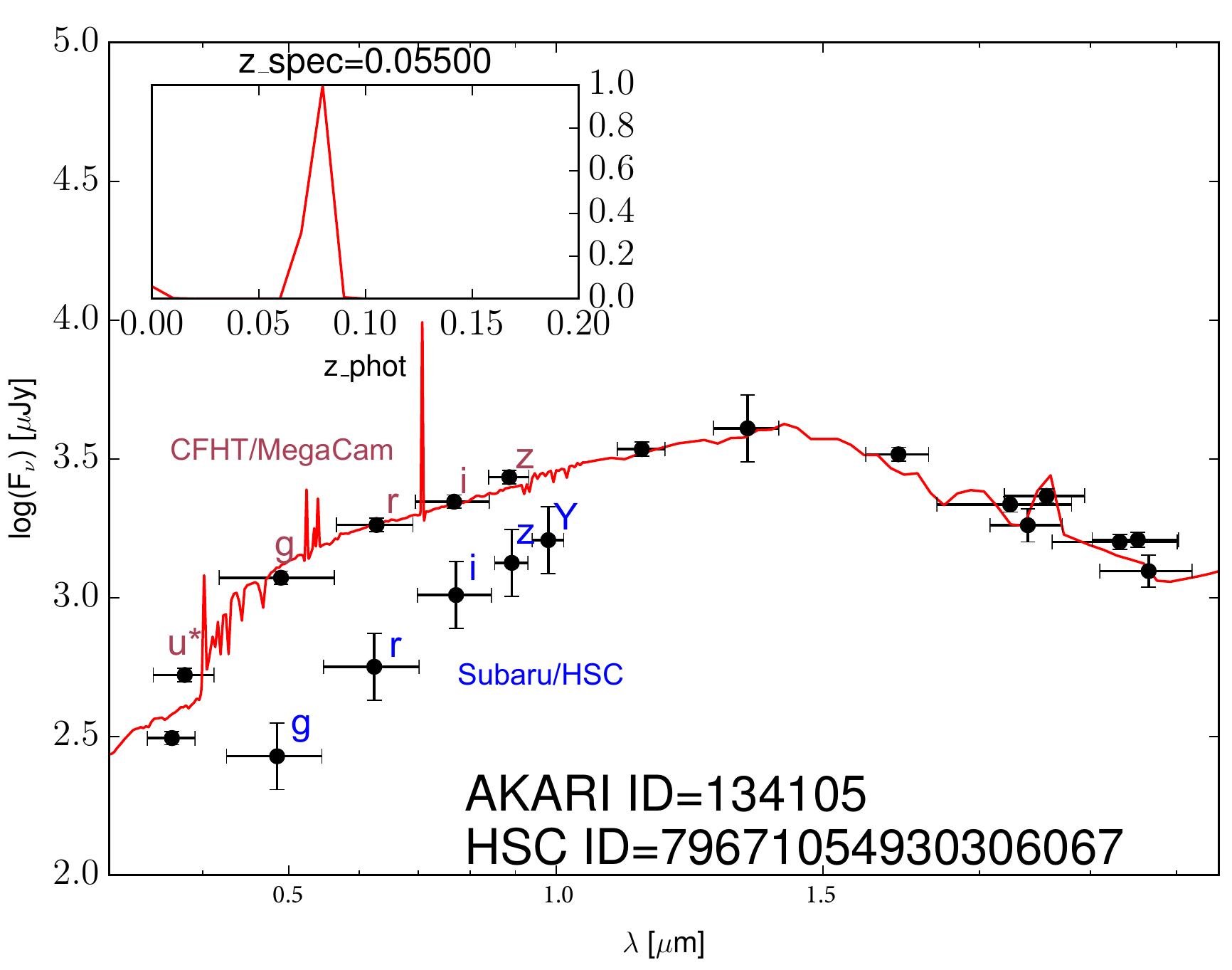}

    \caption{An example SED of a problematic source, where Maidanak/SNUCAM's photometry is inconsistent with Subaru/HSC's. The red curve in the main panel is the best-fitted galaxy SED template. The embedded panel shows the probability distribution function (PDF) of the photo-z.}
    \label{fig:SEDR.pdf}
\end{figure}

To investigate these sources with inconsistent and incorrect magnitudes, we plot SEDs of them. We found that many of these cases are bright sources (MegaCam < 21.0 mag) both in the optical and IR-bands. We show example images in all bands (Fig.~\ref{fig:83827.pdf} ; Note that the images are scaled with different factors to make the structure visible). The red circle in the centre is the position of the $AKARI$ source's coordinates. The cyan circles are Subaru/HSC positions (given by the Subaru/HSC pipeline) near the matching radius. We mark the matched Subaru/HSC source which is cross-matched with the $AKARI$ source in the green circle. Eight of the Subaru/HSC detections are problematic detections. They are mostly parts of the spiral arm of the bright galaxy. Only one of them seems to be a possible deblended source (or part of the HII region of the galaxy) near AKARI\char`_ ID 134105, which appeared at the top left of this spiral galaxy AKARI\char`_ ID 134105 in Subaru/HSC $i$-, $z$-, $Y$-band images. The correct measurement (of the flux) on this galaxy has to be integrated over the entire galaxy. However, the detection and photometry are divided into nine.
This magnitude discrepancy affects the photo-z estimation for certain sources in a significant way. As a problematic source with spec-z, we present the SED of AKARI\char`_ ID = 134105 in Fig.~\ref{fig:SEDR.pdf}. In this case, the CFHT/MegaCam $i$-band magnitude is brighter than Subaru/HSC $i$-band which covers a similar wavelength as the former. As a result, there are two peaks in the probability distribution function (PDF), which is shown in the top-left panel. This indicates that $Le$ $Phare$ is confused by the magnitude discrepancy. In this case, $Le$ $Phare$ tried to take the average of the two magnitudes and fitted the source with a galaxy template of photo-z = 0.0055 (spec-z = 0.028 for this source).

\begin{figure}
\centering
	\includegraphics[width=\columnwidth]{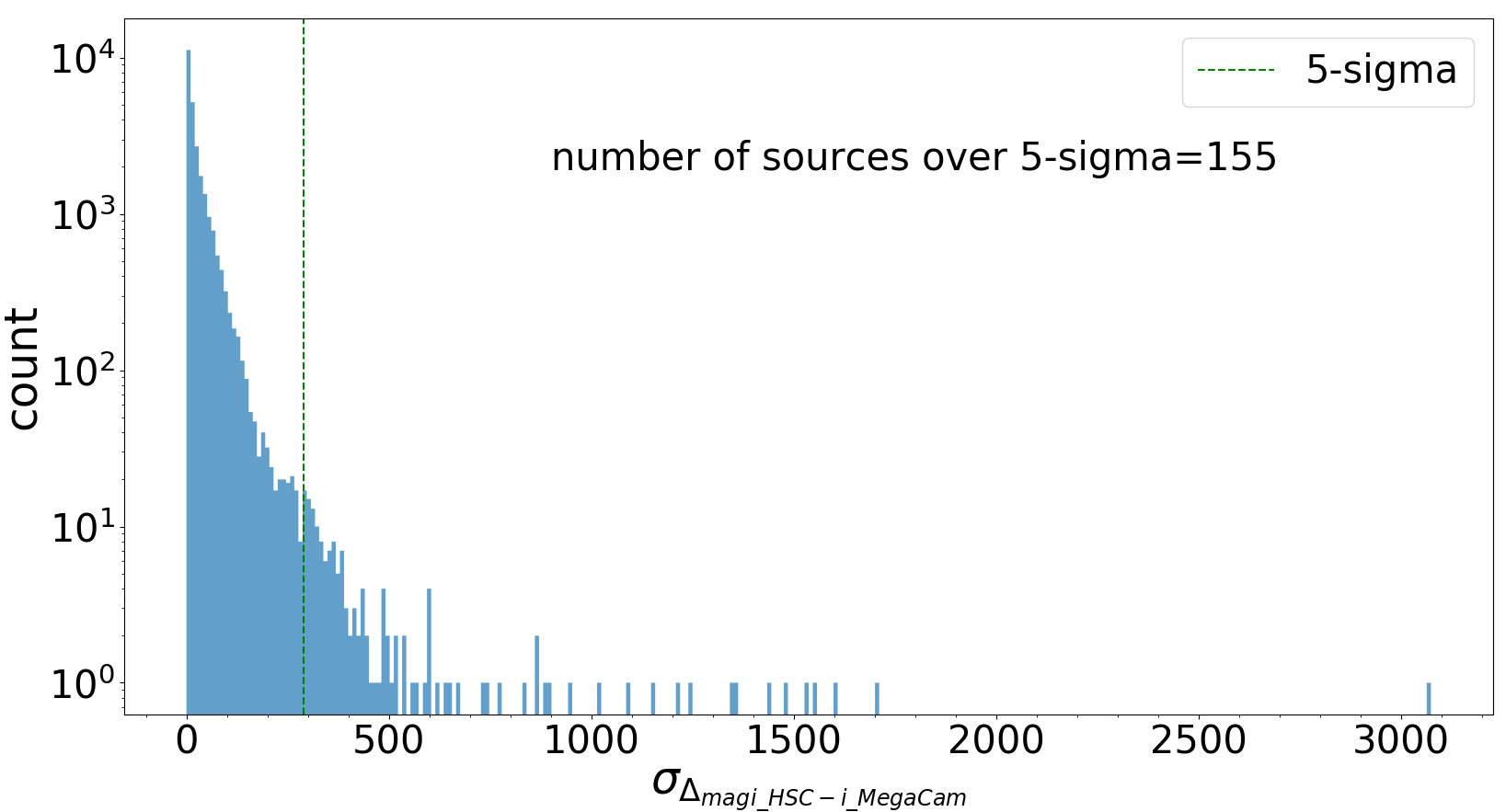}
	\includegraphics[width=\columnwidth]{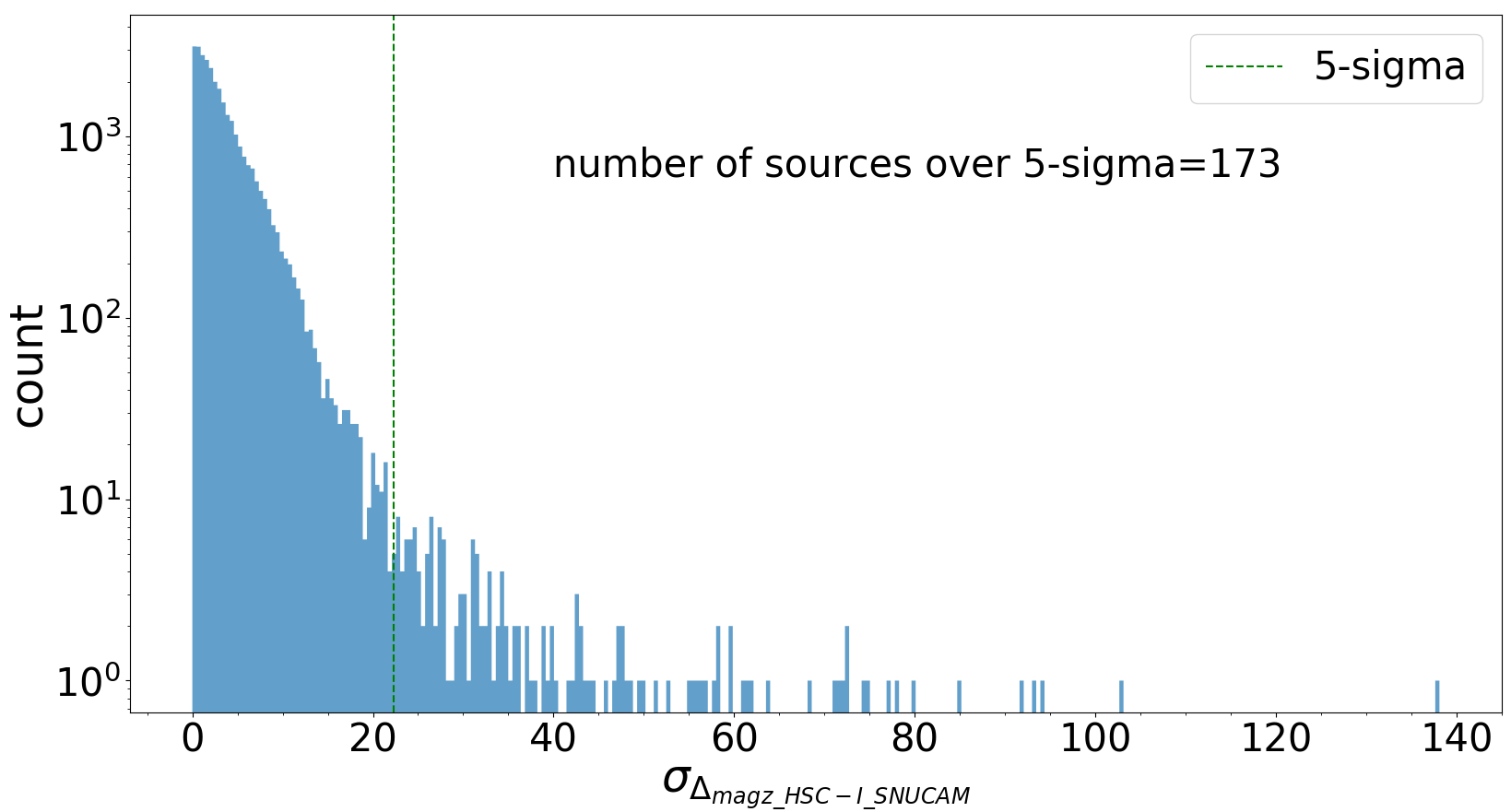}
    \caption{Top panel: histogram of the distribution of
magnitude difference between the Subaru/HSC $i$-band and CFHT/MegaCam $i$-band. 
Bottom panel: histogram of the distribution of
magnitude difference between the Subaru/HSC $z$-band and Maidanak/SNUCAM I-band. The green dashed line in the two panels represents the 5-sigma of the distribution. Note that the y-axis is in logarithmic scale.}
    \label{fig:ii}
\end{figure}

\begin{figure*}
\centering
	\includegraphics[width=110mm]{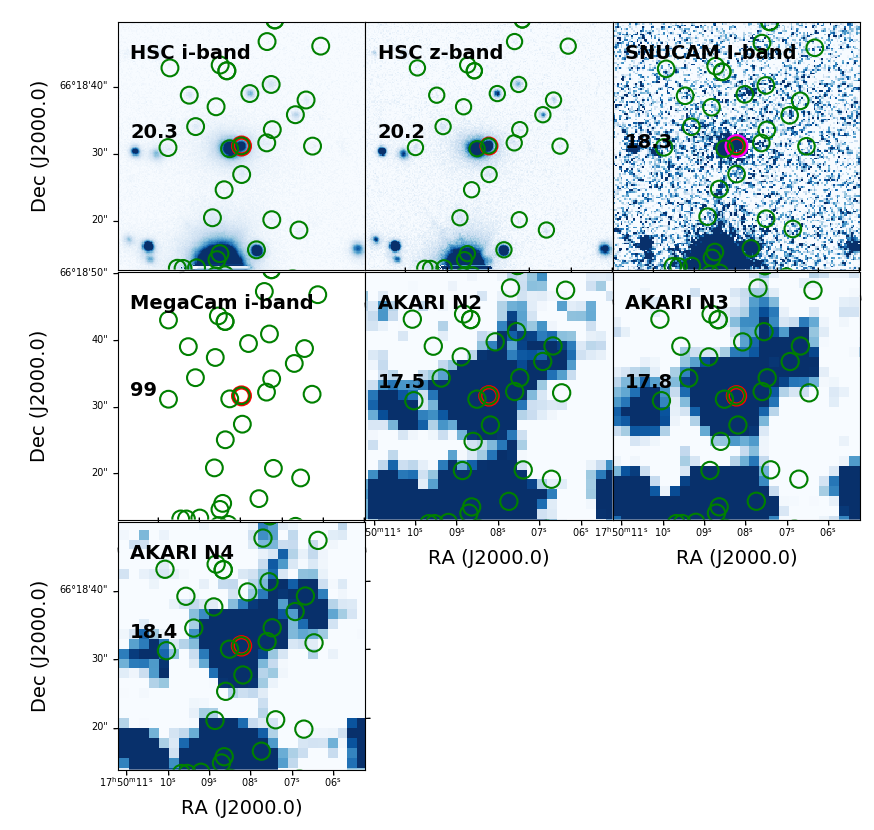}
    \includegraphics[width=110mm]{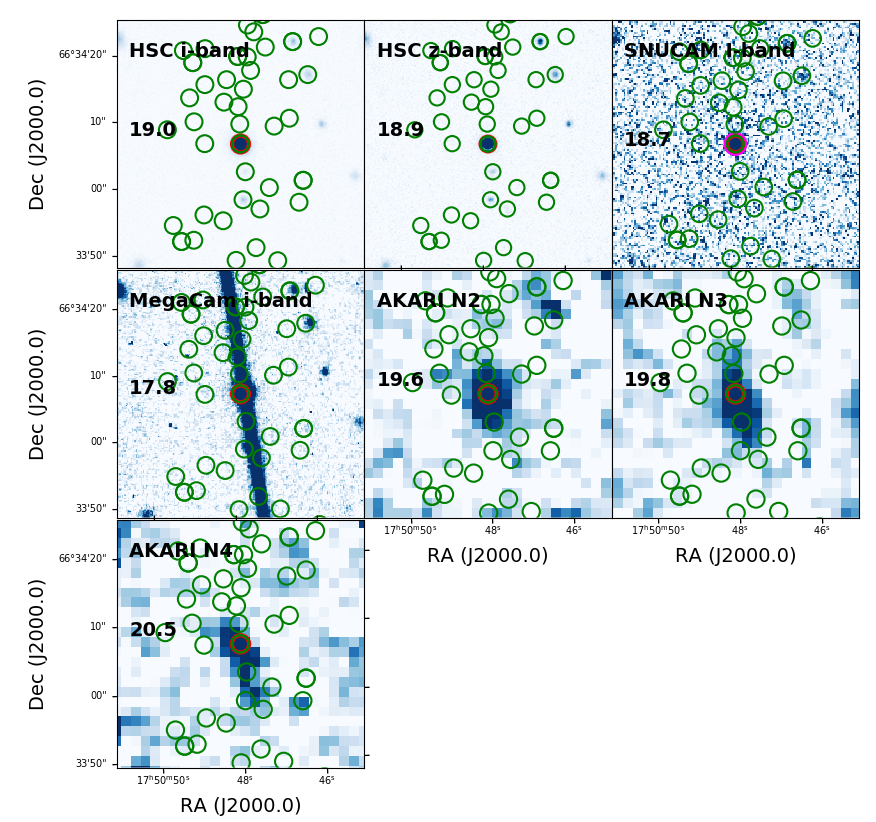}
    
    \caption{Examples of problematic sources mentioned in section \ref{sec:datacal}. The top panel presents a deblended source AKARI\char`_ ID 12072. The bottom panel presents a source AKARI\char`_ ID 55071 with bad photometry in the MegaCam image. The red circle marks the forced coordinate position in HSC. The green circles mark the HSC detection coordinates near the source at forced coordinates. The size of the circles is 2.52 arcsec (definition of aperture size during measurement of photometry).}
    \label{fig:deblend}
\end{figure*}

To prevent poor SED fitting for these problematic sources, we adopt criteria to avoid using 'erroneous' Subaru/HSC photometry for these sources. We calculate the significance of the magnitude difference between the two systems, Subaru/HSC $i$-band ($\lambda_\text{eff}$=7656.0\AA), and CFHT/MegaCam $i$-band ($\lambda_\text{eff}$=7467.4\AA). The significance of the magnitude difference between the two systems is defined as $\sigma_{\Delta_{mag}}=|i_{HSC}-i_{MegaCam}|/\sqrt{\sigma_{i_{HSC}}^{2}+\sigma_{i_{MegaCam}}^{2}}$. We plot the histogram of the significance of the magnitude difference between the two systems and overplot it with its 5$\sigma$. (Fig.~\ref{fig:ii}, top panel). There are 155 out of 26,581 CFHT/MegaCam sources ( < 1 \% ) above the 5$\sigma$ distribution. The number of problematic sources selected by this method is very low. Therefore, applying this criterion does not affect photo-z accuracy significantly. However, it is important to prevent users from using sources with incorrect magnitudes. Therefore, we flag these problematic sources in the online version of the catalogue (Flag\char`_ 2; Table ~\ref{tab:catalog}). 
We found 3 major reasons that caused magnitude discrepancy in the HSC-MegaCam cases: false detections of galaxy structures, deblending of sources, and bad photometry. All cases appear equally often.
We flag these problematic sources (Flag\char`_ 2; Table \ref{tab:catalog}). Great care should be taken if these sources are to be used in research. We present more examples of such problematic source images in Fig. \ref{fig:deblend}.

In some other cases, Subaru/HSC data are inconsistent with Maidanak/SNUCAM data. 
We compare their Subaru/HSC and Maidanak/SNUCAM photometry with a similar way as HSC-MegaCam cases. In this case, Maidanak/SNUCAM $I$-band magnitude is brighter than Subaru/HSC $z$-band which is at a similar wavelength to the former. This kind of source suffers from the same problem as we found in the MegaCam-HSC case. However, there is no SNUCAM-HSC case suffering from bad photometry in the SNUCAM images. The 2 major reasons that caused magnitude discrepancy in HSC-SNUCAM cases: multiple detections of galaxy structures, deblending of sources. Both these problems appear with a similar frequency.

We calculate the significance of the magnitude difference between the two systems, Subaru/HSC $z$-band ($\lambda_\text{eff}$=8902.1\AA), and Maidanak/SNUCAM $I$-band ($\lambda_\text{eff}$=8603.3\AA). The significance of the magnitude difference between the two systems is defined as $\sigma_{\Delta_{mag}}=|z_{HSC}-I_{SNUCAM}|/\sqrt{\sigma_{z_{HSC}}^{2}+\sigma_{I_{SNUCAM}}^{2}}$. We plot the histogram of the significance of the magnitude difference between the two systems and overplot it with its 5$\sigma$. (Fig.~\ref{fig:ii}, bottom panel). There are 173 out of 30,487 Maidanak/SNUCAM sources ( < 1 \% ) above the 5$\sigma$ distribution. We also flag them in the online version of the catalogue (Flag\char`_ 2; Table ~\ref{tab:catalog}).

\begin{table*}
	\centering
	\caption{
	Effective wavelength ($\lambda_\text{eff}$), full width half maximum (FWHM) of the filter transmission, depth, FWHM of point spread function (PSF) and its reference of all the filters we discuss in this paper. A summary of definitions of depths from different sources is presented in APPENDIX \ref{depth}.
	}
	\label{tab:filterinfo}
	\begin{tabular}{ccccccc} % four columns, alignment for each
		\hline\hline 
		Filter & $\lambda_\text{eff}$ $($\AA$)$ & FWHM $($\AA$)$ & 5$\sigma$ depth mag [$\mu$Jy] & FWHM of PSF (") & Reference \\
		\hline
		Maidanak/SNUCAM-$B$ & 4242.9 & 926.5 & 23.4 [1.58] & 1.40 & \citet{Jeon2010} \\
		Maidanak/SNUCAM-$R$ & 5402.3 & 1459.7 & 23.1 [2.09] & 1.20 & \citet{Jeon2010} \\
		Maidanak/SNUCAM-$I$ & 8603.3 & 3370.2 & 22.3 [4.36] & 1.10 & \citet{Jeon2010} \\
		Subaru/HSC-$g$ & 4723.4 & 1386.6 & 28.6 [0.01] & 0.68 & Oi et al. 2020 (in press) \\
		Subaru/HSC-$r$ & 6136.2 & 1502.9 & 27.3 [0.04] & 1.25 & Oi et al. 2020 (in press) \\
		Subaru/HSC-$i$ & 7656.0 & 1553.6 & 26.7 [0.08] & 0.84 & Oi et al. 2020 (in press) \\
		Subaru/HSC-$z$ & 8902.1 & 773.0 & 26.0 [0.14] & 0.76 & Oi et al. 2020 (in press) \\
		Subaru/HSC-$Y$ & 9749.8 & 782.8 & 25.6 [0.21] & 0.73 & Oi et al. 2020 (in press) \\
		CFHT/MegaPrime-$u$ & 3827.2 & 758.4 & 25.8 [1.04] & 0.94 & \citet{Huang2020} \\
		CFHT/MegaCam-$u^{*}$ & 3881.6 & 653.7 & 25.9 [0.16] & 0.94 & \citet{Oi2014} \\ 
		CFHT/MegaCam-$g$ & 4767.0 & 1433.8 & 26.1 [0.13] & 0.96 & \citet{Oi2014} \\ 
		CFHT/MegaCam-$r$ & 6191.7 & 1218.8 & 25.6 [0.21] & 0.88 & \citet{Oi2014} \\
		CFHT/MegaCam-$i$ & 7467.4 & 1570.9 & 24.9 [0.39] & 0.84 & \citet{Oi2014} \\
		CFHT/MegaCam-$z$ & 8824.0 & 935.5 & 24.0 [0.91] & 0.86 & \citet{Oi2014} \\
		KPNO/FLAMINGOS-$J$ & 12407.9 & 1563.5 & 21.6 [8.30] & 1.80 & \citet{Jeon2014} \\
		KPNO/FLAMINGOS-$H$ & 16221.3 & 2807.8 & 21.3 [11.0] & 1.70 & \citet{Jeon2014} \\
		CFHT/WIRCam-$Y$ & 10258.8 & 1105.3 & 23.4 [1.58] & 0.80 & \citet{Oi2014} \\
		CFHT/WIRCam-$J$ & 12481.5 & 1587.5 & 23.0 [2.29] & 0.81 & \citet{Oi2014} \\
		CFHT/WIRCam-$K_s$ & 21337.8 & 3270.5 & 22.7 [3.02] & 0.66 & \citet{Oi2014} \\
		$AKARI$/IRC-$N2$ & 22684.4 & 8468.7 & 20.9 [15.4] & 4.80 & \citet{Kim2012} \\
		$AKARI$/IRC-$N3$ & 31302.7 & 11573.9 & 21.1 [13.3] & 4.80 & \citet{Kim2012} \\
		$AKARI$/IRC-$N4$ & 42505.2 & 16343.2 & 21.1 [13.6] & 4.90 & \citet{Kim2012} \\		
		WISE1 & 33156.6 & 6357.9 & 18.1 [18.0] & 6.10 & \citet{Jarrett2011} \\ %https://www.cfa.harvard.edu/irac/
		WISE2 & 45645.0 & 11073.2 & 17.2 [23.0] & 6.40 & \citet{Jarrett2011} \\
		$Spitzer$/IRAC1 & 35075.1 & 7431.7 & 21.8 [6.45] & 1.97 & \citet{Nayyeri2018} \\
		$Spitzer$/IRAC2 & 44365.8 & 10096.8 & 22.4 [3.95] & 1.93 & \citet{Nayyeri2018} \\
		\hline
	\end{tabular}
\end{table*}

\begin{landscape}
\begin{table}
	\caption{Photo-z catalogue at $AKARI$ NEPW field.}
	\begin{center} 
	\label{tab:catalog}
	\setlength{\tabcolsep}{2pt}
	\begin{tabular}{|*{26}{c|}} % four columns, alignment for each
		\hline
		\hline
		AKARI\char`_ID & HSC\char`_ ID & Tri\char`_ Ptch & forced\char`_ RA & forced\char`_Dec & photo-z & z\char`_ 68\char`_ low & z\char`_ 68\char`_ high & spec-z & weight\\
		\hline
		40407 & 79217643822780069 & 18012\char`_ 0\char`_ 7 & 270.7276917 & 65.3166046 & 0.5612 & 0.4820 & 0.6329 & -99.000 & -99.000\\       
        101696 & 79217643822780251 & 18012\char`_ 0\char`_ 7 & 270.7315979 & 65.3317566 & 1.8333 & 1.4818 & 2.6538 & -99.000 & -99.000\\       
        101445 & 79217643822780303 & 18012\char`_ 0\char`_ 7 & 270.7599487 & 65.3283157 & 0.3794 & 0.3682 & 0.3955 & -99.000 & -99.000\\       
        104332 & 79217643822780314 & 18012\char`_ 0\char`_ 7 & 270.7807617 & 65.3384857 & 0.3930 & 0.2442 & 0.4149 & -99.000 & -99.000\\       
        126629 & 79217643822780319 & 18012\char`_ 0\char`_ 7 & 270.7976379 & 65.3421173 & 0.3468 & 0.3004 & 0.3985 & -99.000 & -99.000\\       
        102997 & 79217643822780420 & 18012\char`_ 0\char`_ 7 & 270.7579346 & 65.3350601 & 0.3762 & 0.3602 & 0.3921 & -99.000 & -99.000\\       
        99701 & 79217643822780435 & 18012\char`_ 0\char`_ 7 & 270.7673035 & 65.3377075 & 0.3835 & 0.3622 & 0.4083 & -99.000 & -99.000\\       
        48446 & 79217643822780448 & 18012\char`_ 0\char`_ 7 & 270.7470093 & 65.3519974 & 0.3777 & 0.3546 & 0.4003 & -99.000 & -99.000\\       
        129078 & 79217643822780459 & 18012\char`_ 0\char`_ 7 & 270.7839050 & 65.3499985 & 0.6312 & 0.5254 & 0.6779 & -99.000 & -99.000\\       
        104610 & 79217643822780460 & 18012\char`_ 0\char`_ 7 & 270.7848206 & 65.3462601 & 1.1936 & 1.0663 & 1.2446 & -99.000 & -99.000\\       
        129036 & 79217643822780559 & 18012\char`_ 0\char`_ 7 & 270.7474670 & 65.3552322 & 0.7245 & 0.6596 & 0.7973 & -99.000 & -99.000\\       
        104562 & 79217643822780742 & 18012\char`_ 0\char`_ 7 & 270.7695312 & 65.3519974 & 1.1540 & 1.0483 & 1.2614 & -99.000 & -99.000\\       
        88044 & 79217643822780857 & 18012\char`_ 0\char`_ 7 & 270.7791748 & 65.3472900 & 2.4175 & 1.8768 & 2.8525 & -99.000 & -99.000\\   
        113377 & 79217643822780887 & 18012\char`_ 0\char`_ 7 & 270.7696533 & 65.3484039 & 1.1419 & 1.0452 & 1.2574 & -99.000 & -99.000\\  
        1473 & 79217643822780937 & 18012\char`_ 0\char`_ 7 & 270.7780762 & 65.3443909 & 1.2626 & 1.0450 & 2.2645 & -99.000 & -99.000\\
        \\
        \hline
		\hline
		Flag\char`_ 1 & Flag\char`_ 2 & chi\char`_ gal & chi\char`_ star & chi\char`_ qso & PDF\\
		\hline
		0 & 0 & 19.6707 & 75.9512 & 12.7039 & 96.256 \\
		0 & 0 & 18.3821 & 35.0927 & 18.3848 & 64.284 \\
		1 & 0 & 131.217 & 26.9841 & 166.743 & 100.000 \\
		1 & 0 & 165.291 & 20.9277 & 190.292 & 72.310 \\
		1 & 1 & 51.1234 & 9.81378 & 63.0844 & 96.806 \\
		1 & 0 & 90.1393 & 16.6556 & 114.741 & 100.000 \\
		1 & 0 & 64.3464 & 12.1754 & 80.6374 & 99.740 \\
		0 & 0 & 7.94405 & 96.9727 & 7.97329 & 99.905 \\
		0 & 0 & 19.6331 & 65.5227 & 28.141 & 95.592 \\
		0 & 0 & 18.5843 & 76.4993 & 22.917 & 91.188 \\
		0 & 0 & 9.52094 & 170.119 & 12.3986 & 87.933 \\
		0 & 0 & 3.41178 & 84.7332 & 5.72297 & 93.921 \\
		0 & 0 & 3.58921 & 39.5975 & 3.41148 & 49.679 \\
		0 & 0 & 6.40542 & 63.7636 & 8.51876 & 88.557 \\
		0 & 0 & 3.44284 & 25.6715 & 2.6416 & 45.336 \\
		\hline
\multicolumn{2}{l}{Note: The full catalogue of this table is available on reasonable request to the corresponding author.}		

	\end{tabular}
	\end{center}
\end{table}
\end{landscape}

 \begin{table*}
	\centering
	\caption{Different photo-z results with different SED templates at photo-z < 1.5. The weighted photo-z performance is shown in square brackets.}
	\label{tab:models}
	\begin{tabular}{ccccccc} % four columns, alignment for each
		\hline
		Set & Template name & N$_{gal}$ & $\eta$ [weighted $\eta$] & {$\sigma_{\Delta{z/(1+z)}}$} [weighted $\sigma_{\Delta{z/(1+z)}}$]\\
		\hline
		 A1 & COSMOS & 1,921 & 13.3\% [11.3\%] & 0.054 [0.053] \\ 
		 A2 & SALVATO & 1,906 & 22.9\% [29.9\%] & 0.105 [0.130] \\
		 A3 & SWIRE & 1,949 & 19.6\% [17.6\%] & 0.085 [0.085] \\
		 A4 & AVEROIN\char`_NEW & 1,942 & 16.3\% [14.8\%] & 0.066 [0.062] \\
		 A5 & BROWN & 1,937 & 17.0\% [16.2\%] & 0.060 [0.060] \\
		 A6 & CWW & 1,921 & 17.9\% [17.8\%] & 0.092 [0.101] \\
		 A7 & COSMOS\char`_AGN & 1,941 & 16.5\% [17.2\%] & 0.056 [0.055] \\
		\hline
	\end{tabular}
\end{table*}

\section{Photo-z with SED template fitting}\label{sec:SED}
We derived the photo-z of the NEPW sources using the software $Le$ $Phare$ \citep{Arnouts1999, Ilbert2006}, which is based on a $\chi^2$ template fitting method. We use a function, \texttt{AUTO\_ADAPT}, in $Le$ $Phare$ for adjusting magnitude zero-points to match the templates with the existing spec-z. Typical magnitude offsets from \texttt{AUTO\_ADAPT} are $\sim$0.02 for most of the tests throughout the paper. We also slightly adjust the error by adding 0.05 mag to Subaru/HSC photometry in the quadrature to compute better photo-z following \citet{Aihara2018}, and Oi et al. 2020 (in press). They have suggested this error to avoid underestimation of HSC magnitude uncertainty that may lead to an unrealistic PDF.

We assessed the photo-z accuracy by comparing with the spec-z. We can estimate the photo-z dispersion from {$\sigma_{\Delta{z/(1+z)}}$} which is conventionally defined using the normalised median absolute deviation \citep[NMAD(|$\Delta$z|);][]{Hoaglin1983} with {$\sigma_{\Delta{z/(1+z)}}$} =1.48$\times$NMAD(|$\Delta$z|). This dispersion estimate is robust to catastrophic error rate, $\eta$ and outliers are defined as |$\Delta$z|/(1 + $z_s$)> 0.15. These are the two main indicators of photo-z performance. A smaller photo-z dispersion and catastrophic error rate indicate a better photo-z performance. We define N$_\text{gal}$ as a number of galaxies with photo-z computed in the test within certain redshift.

Photo-z estimation is a regression type of analysis. Therefore, the process of learning the relationships among variables is very important. In this paper, we focus on two major parameters, which are the template model, and filter combination. We compare different parameter set-ups and find out the best set-up in our NEP band-merged catalogue case.

\subsection{Comparison of different template models}\label{sec:model}
SED template fitting involves building a library of observed and/or model templates for many redshifts and matching these to the observed SEDs to estimate the redshift. Therefore, choosing the best template model is very important.

We compare different photo-z results by 6 different commonly used template libraries: COSMOS \citep{Ilbert2009}, SWIRE \citep{Polletta2007}, AVEROIN\char`_ NEW \citep{Arnouts1999}, BROWN \citep{Brown2014}, CWW \citep{Coleman1980, Calzetti1994}, SALVATO \citep{Salvato2009}. There is no active galactic nuclei (AGN) template in the COSMOS template library. Therefore, we add 6 AGN templates from SWIRE and make a new template library. We call this COSMOS\char`_ AGN. Our controlled variable is 26-bands filter combination, which are Maidanak/SNUCAM $B$, $R$, $I$, Subaru/HSC $g$,  $r$, $i$, $z$, $Y$, CFHT/MegaCam $u^{*}$, $g$, $r$, $i$, $z$, CFHT/MegaPrime $u$, KPNO/FLAMINGOS $H$, CFHT/WIRCam $Y$, $J$, $K_s$, $AKARI$/IRC $N2$, $N$3, $N4$, WISE W1, W2, and $Spitzer$/IRAC 1 and 2. Our target redshift range is 0 < z < 1.5.
We summarise the comparison of results obtained using different template libraries in Table ~\ref{tab:models}. We use the median as a point estimate for estimation of the photo-z. Note that some models failed to compute photo-z for some of the sources due to the large reduced $\chi^2$ value (>1,500) when the photometry is fitted to the galaxy templates. There are only 3 out of 2026 sources suffering from this problem in the spectroscopic sample. %Therefore, the photo-z becomes undefined and the number of sources with photo-z (N$_{gal}$) decreases.

After we compare different template models, we find that COSMOS performs the best in terms of both catastrophic error and photo-z dispersion for our band-merged catalogue. It achieves a redshift dispersion of {$\sigma_{\Delta{z/(1+z)}}$} = 0.054 [0.053] with the catastrophic error of $\eta$ =13.3\% [11.3\%] at z < 1.5. The weighted photo-z performance is shown in square brackets. In COSMOS, there are 31 model templates for different types of galaxies and it is also the library with the highest number of templates. This is very likely to be the reason why COSMOS performs the best in this case.

\subsection{Comparison of different filter combinations}\label{sec:filtercomb}
Optical bands are important, as low-redshift galaxies' spectral break features in the SED are significant at those wavelengths. Moreover, the $u^{*}$-band is important because they can trace the Balmer break feature in an SED at very low redshift. NIR data are also precious in our analysis as it can trace the Balmer break feature in an SED out to higher redshift. Here, we compare different combinations among 26-band optical-NIR photometric data. We investigate which combination is most suitable for our photo-z analysis. For each band combination, we performed an SED fitting analysis to derive photo-z accuracy. We summarise the comparison of the results obtained with different filters combinations in Table~\ref{tab:filtercomb}.

After we compare different template models, we find that set B1, 26 bands photometry performs the best in terms of both catastrophic error and photo-z dispersion for our band-merged catalogue. They are Maidanak/SNUCAM $B$, $R$, $I$, Subaru/HSC $g$, $r$, $i$, $z$, $Y$, CFHT/MegaCam $u^{*}$, $g$, $r$, $i$, $z$, CFHT/MegaPrime $u$, KPNO/FLAMINGOS $J$, $H$, CFHT/WIRCam $Y$, $J$, $K_s$, $AKARI$/IRC $N2$, $N3$, $N4$, WISE W1, W2, and $Spitzer$/IRAC 1 and 2. It achieves a redshift dispersion of {$\sigma_{\Delta{z/(1+z)}}$} = 0.054 [0.053] with the catastrophic error of $\eta$ =13.3\% [11.3\%]. The weighted photo-z performance is shown in square brackets. Note that set filter combination B7 performs worse than B1. The only difference between these two is the absence of $AKARI$/IRC $N2$, $N3$, $N4$-bands. This emphasises the strength of the NEPW survey is the incorporation of $AKARI$ data which improves photo-z performance. Set B12 is the filter combination used by \citet{Oi2014}, who only analyzed the NEPD, a smaller region inside the NEPW field. As there are just a few filter sets, the number of sources decreases significantly. This indicates the importance of other filters for the sources outside the deep field region.

\begin{table*}
	\centering
	\caption{Different photo-z results by different filter combinations for photo-z < 1.5. The weighted photo-z performance is shown in square brackets.}
	\label{tab:filtercomb}
	\resizebox{\textwidth}{!}{
	\begin{tabular}{llllll} % four columns, alignment for each
        \hline
        Set & No. of filters & N$_\text{gal}$ & $\eta$ [weighted $\eta$] & {$\sigma_{\Delta{z/(1+z)}}$} [weighted $\sigma_{\Delta{z/(1+z)}}$] & Filter combination\\
        \hline
        B1 & 26 & 1,924 & 13.3\% [11.3\%] & 0.054 [0.053] &  Maidanak/SNUCAM $B$, $R$, $I$, Subaru/HSC $g$,  $r$, $i$, $z$, $Y$,\\
        &&&&& CFHT/MegaCam $u^{*}$, $g$, $r$, $i$, $z$, CFHT/MegaPrime $u$,\\
        &&&&& KPNO/FLAMINGOS $J$, $H$, CFHT/WIRCam $Y$, $J$, $K_s$,\\
        &&&&& $AKARI$ $N2$, $N3$, $N4$, $Spitzer$/IRAC1, 2, WISE 1, 2\\
        \hline
        B2 & 23 & 1,925 & 17.1\% [14.5\%] & 0.077 [0.066] & Subaru/HSC $g$, $r$, $i$, $z$, $Y$, CFHT/MegaCam $u^{*}$, $g$, $r$, $i$, $z$,\\ 
        &&&&& CFHT/MegaPrime $u$, KPNO/FLAMINGOS $J$, $H$,\\ 
        &&&&& CFHT/WIRCam $Y$, $J$, $K_s$, $AKARI$ $N2$, $N3$, $N4$,\\ 
        &&&&& $Spitzer$/IRAC1, 2, WISE 1, 2\\
        \hline
        B3 & 21 & 1,933 & 15.5\% [15.4\%] & 0.060 [0.066] &  Maidanak/SNUCAM $B$, $R$, $I$, CFHT/MegaCam $u^{*}$, $g$, $r$, $i$, $z$, \\
        &&&&& CFHT/MegaPrime $u$, KPNO/FLAMINGOS $J$, $H$,\\ 
        &&&&& CFHT/WIRCam $Y$, $J$, $K_s$, $AKARI$ $N2$, $N3$, $N4$,\\ 
        &&&&& $Spitzer$/IRAC1, 2, WISE 1, 2\\
        \hline
        B4 & 21 & 1,908 & 14.9\% [14.9\%] & 0.067 [0.064] &  Maidanak/SNUCAM $B$, $R$, $I$, Subaru/HSC $g$, $r$, $i$, $z$, $Y$,\\
        &&&&& KPNO/FLAMINGOS $J$, $H$, CFHT/WIRCam $Y$, $J$, $K_s$,\\
        &&&&& $AKARI$ $N2$, $N3$, $N4$,\\
        &&&&& $Spitzer$/IRAC1, 2, WISE 1, 2\\		  
        \hline
        B5 & 24 & 1931 & 12.6\% [11.7\%] & 0.053 [0.053] &  Maidanak/SNUCAM $B$, $R$, $I$, Subaru/HSC $g$, $r$, $i$, $z$, $Y$,\\
        &&&&& CFHT/MegaCam $u^{*}$, $g$, $r$, $i$, $z$, CFHT/MegaPrime $u$,\\
        &&&&& CFHT/WIRCam $Y$, $J$, $K_s$, $AKARI$ $N2$, $N3$, $N4$,\\
        &&&&& $Spitzer$/IRAC1, 2, WISE 1, 2\\
        \hline
        B6 & 23 & 1917 & 16.3\% [13.0\%] & 0.057 [0.062] &  Maidanak/SNUCAM $B$, $R$, $I$, Subaru/HSC $g$,  $r$, $i$, $z$, $Y$,\\
        &&&&& CFHT/MegaCam $u^{*}$, $g$, $r$, $i$, $z$, CFHT/MegaPrime $u$,\\
        &&&&& $AKARI$ $N2$, $N3$, $N4$, KPNO/FLAMINGOS $J$, $H$, \\
        &&&&& $Spitzer$/IRAC1, 2, WISE 1, 2\\
        \hline
        B7 & 23 & 1,928 & 14.2\% [13.2\%] & 0.057 [0.054] &  Maidanak/SNUCAM $B$, $R$, $I$, Subaru/HSC $g$,  $r$, $i$, $z$, $Y$,\\
        &&&&& CFHT/MegaCam $u^{*}$, $g$, $r$, $i$, $z$, CFHT/MegaPrime $u$,\\
        &&&&& KPNO/FLAMINGOS $J$, $H$, CFHT/WIRCam $Y$, $J$, $K_s$,\\
        &&&&& $Spitzer$/IRAC1, 2, WISE 1, 2\\
        \hline
        B8 & 22 & 1,928 & 15.3\% [13.3\%] & 0.060 [0.057] &  Maidanak/SNUCAM $B$, $R$, $I$, Subaru/HSC $g$, $r$, $i$, $z$, $Y$,\\
        &&&&& CFHT/MegaCam $u^{*}$, $g$, $r$, $i$, $z$, CFHT/MegaPrime $u$,\\
        &&&&& KPNO/FLAMINGOS $J$, $H$, CFHT/WIRCam $Y$, $J$, $K_s$,\\
        &&&&& $AKARI$ $N2$, $N3$, $N4$,\\
        \hline
        B9 & 24 & 1,921 & 13.7\% [12.6\%] & 0.054 [0.054] &  Maidanak/SNUCAM $B$, $R$, $I$, Subaru/HSC $g$, $r$, $i$, $z$, $Y$,\\
        &&&&& CFHT/MegaCam $u^{*}$, $g$, $r$, $i$, $z$, CFHT/MegaPrime $u$,\\
        &&&&& KPNO/FLAMINGOS $J$, $H$, CFHT/WIRCam $Y$, $J$, $K_s$,\\
        &&&&& $AKARI$ $N2$, $N3$, $N4$, $Spitzer$/IRAC1, 2\\
		\hline
	    B10 & 24 & 1,930 & 14.7\% [13.1\%] & 0.056 [0.056] &  Maidanak/SNUCAM $B$, $R$, $I$, Subaru/HSC $g$, $r$, $i$, $z$, $Y$,\\
        &&&&& CFHT/MegaCam $u^{*}$, $g$, $r$, $i$, $z$, CFHT/MegaPrime $u$,\\
        &&&&& KPNO/FLAMINGOS $J$, $H$, CFHT/WIRCam $Y$, $J$, $K_s$,\\
        &&&&& $AKARI$ $N2$, $N3$, $N4$, WISE 1, 2\\
		\hline
	    B11 & 25 & 1917 & 14.2\% [13.6\%] & 0.071 [0.063] &  Maidanak/SNUCAM $B$, $R$, $I$, Subaru/HSC $g$, $r$, $i$, $z$, $Y$,\\
        &&&&& CFHT/MegaCam $g$, $r$, $i$, $z$, CFHT/MegaPrime $u$,\\
        &&&&& KPNO/FLAMINGOS $J$, $H$, CFHT/WIRCam $Y$, $J$, $K_s$,\\
        &&&&& $AKARI$ $N2$, $N3$, $N4$, $Spitzer$/IRAC1, 2, WISE 1, 2\\
		\hline
	    B12 & 8 & 1092 & 14.2\% [13.6\%] & 0.063 [0.064] &  
        CFHT/MegaCam $u^{*}$, $g$, $r$, $i$, $z$, 
        CFHT/WIRCam $Y$, $J$, $K_s$,\\
		\hline
	\end{tabular}}
\end{table*}

\subsection{Star/Galaxy Classification}
For each source, $\chi^2$ value is evaluated for both the galaxy templates and stellar SED templates \citep{Pickles1998, Chabrier2000, Bohlin1995}. We use the $\chi^2$ value to separate stars and galaxies. Note that we rely on the SED when separating stars and galaxies, rather than on morphological information such as stellarity and extendedness. This is because we do not want to exclude AGN/quasars as point sources. If $\chi^2_{gal} > \chi^2_{star}$, where $\chi^2_{gal}$ and $\chi^2_{star}$ are the minimum $\chi^2$ values obtained with the galaxy and stellar templates, respectively, the object is flagged as a possible star.

We use spectroscopically confirmed samples \citep{Krumpe2015} to examine the performance of the SED $\chi^2$ classification. For stars, they were in excellent agreement: 5/5 of the spectroscopically confirmed stars are classified correctly. For galaxies, 1,588/1,616 of the spectroscopically confirmed galaxies are classified correctly. Note that we are not focusing on AGN classification in this paper, but only separating the stars from this catalogue samples. AGN classification of the NEPW field will be discussed in \citet{Wang2020}.

\begin{figure*}
    \includegraphics[width=140mm]{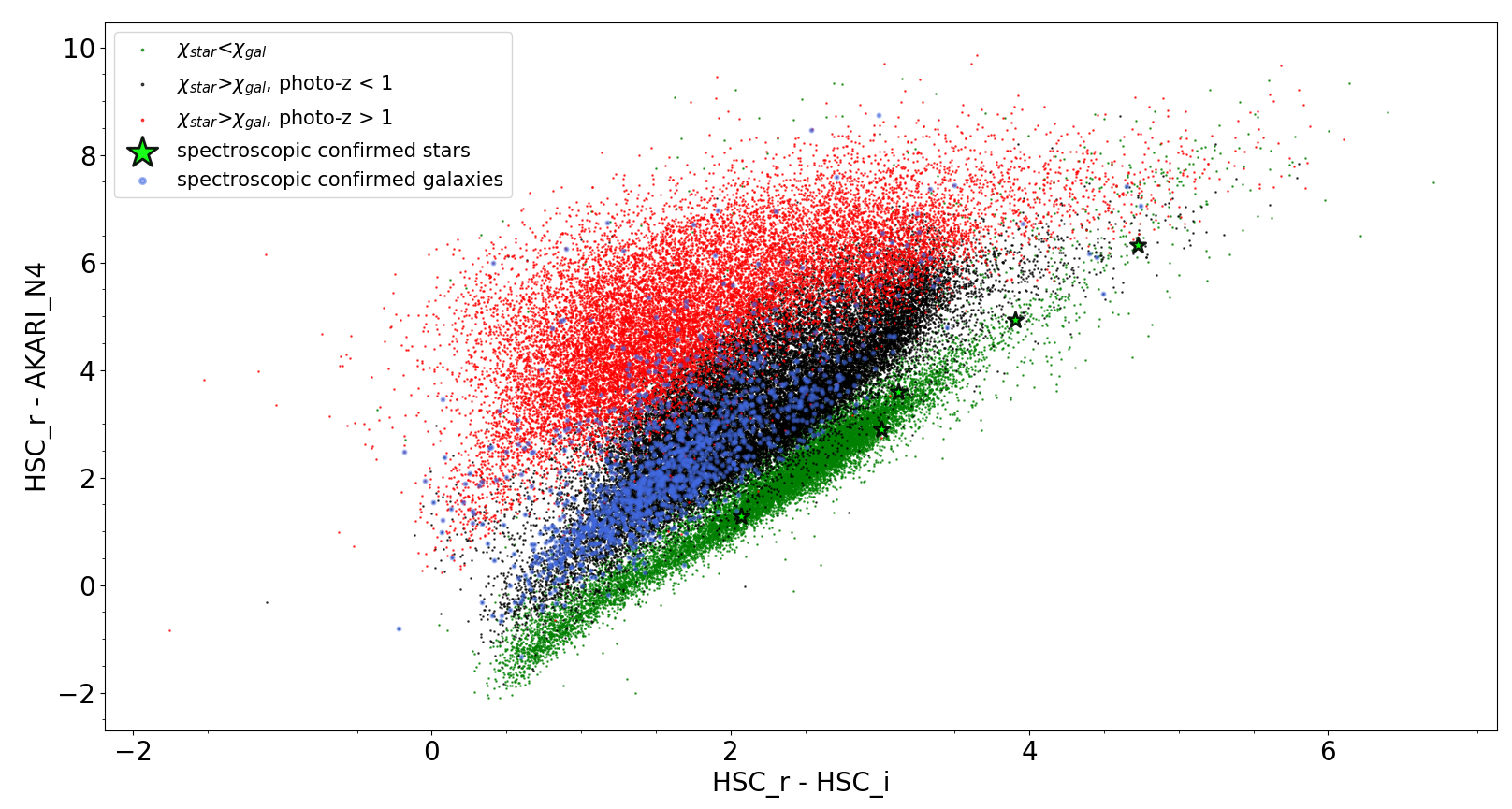}
    \caption{Colour-colour diagnostics of
    Subaru/HSC $g$-band minus $AKARI$ $N4$ versus Subaru/HSC $r$-band minus Subaru/HSC $i$-band. Red dots and black dots are sources with $\chi^2_{gal}$ < $\chi^2_{star}$, redshift > 1 and redshift < 1, respectively. Green dots are sources with  $\chi^2_{star}$ < $\chi^2_{gal}$. Green star-symbols with black edge are spectroscopically confirmed stars. Blue dots are spectroscopically confirmed galaxies.}
    \label{fig:morestardiag.png}
\end{figure*}

After separating stars and galaxies, we use colour-colour diagnostics to check the performance of our star/galaxy classification \citep[e.g.,][]{Ilbert2009, Daddi2004}. We plot Subaru/HSC $g$-band minus $AKARI$ $N4$ versus Subaru/HSC $g$-band minus Subaru/HSC $i$-band. Fig. \ref{fig:morestardiag.png} shows 3 different sequences, which are well distinguished by their colours. Sources marked in red and black are sources with $\chi^2_{gal}$ < $\chi^2_{star}$, which are classified as galaxies with z > 1 and z < 1, respectively. Sources in green are sources with  $\chi^2_{star}$ < $\chi^2_{gal}$, which are classified as stars. Green stellar sequence is isolated at the bottom of the three sequences. Spectroscopically confirmed stars and galaxies are also plotted on Fig. \ref{fig:morestardiag.png}. Blue dots and green stars are evenly distributed on the red+black and green dots, respectively.
There are 24,110 out of 91,861 objects with $\chi^2_{gal}$ > $\chi^2_{star}$. They are flagged as possible stars in this NEP photo-z catalogue (Flag\char`_ 1; Table \ref{tab:catalog}).

\subsection{Best parameters for photo-z accuracy}
\begin{figure}
\centering
	\includegraphics[width=\linewidth]{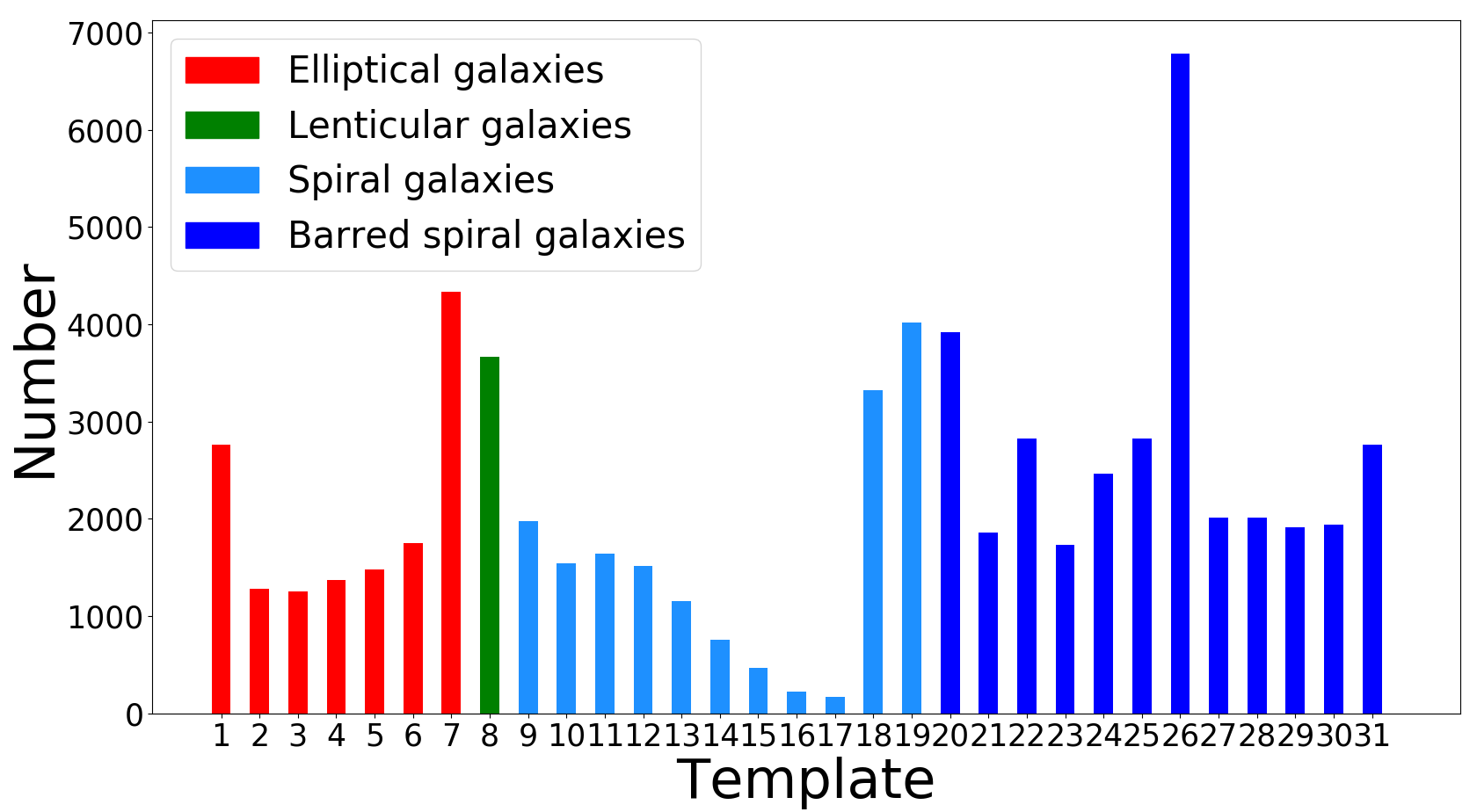}

    \caption{The histogram of different templates used when running photo-z with the whole catalogue. The template information and spectra is listed in Table \ref{tab:COSMOS} and Fig.\ref{fig:tempspectra}, respectively.}
    \label{fig:templatehist.pdf}
\end{figure}

\begin{figure}
\centering
	\includegraphics[width=\columnwidth]{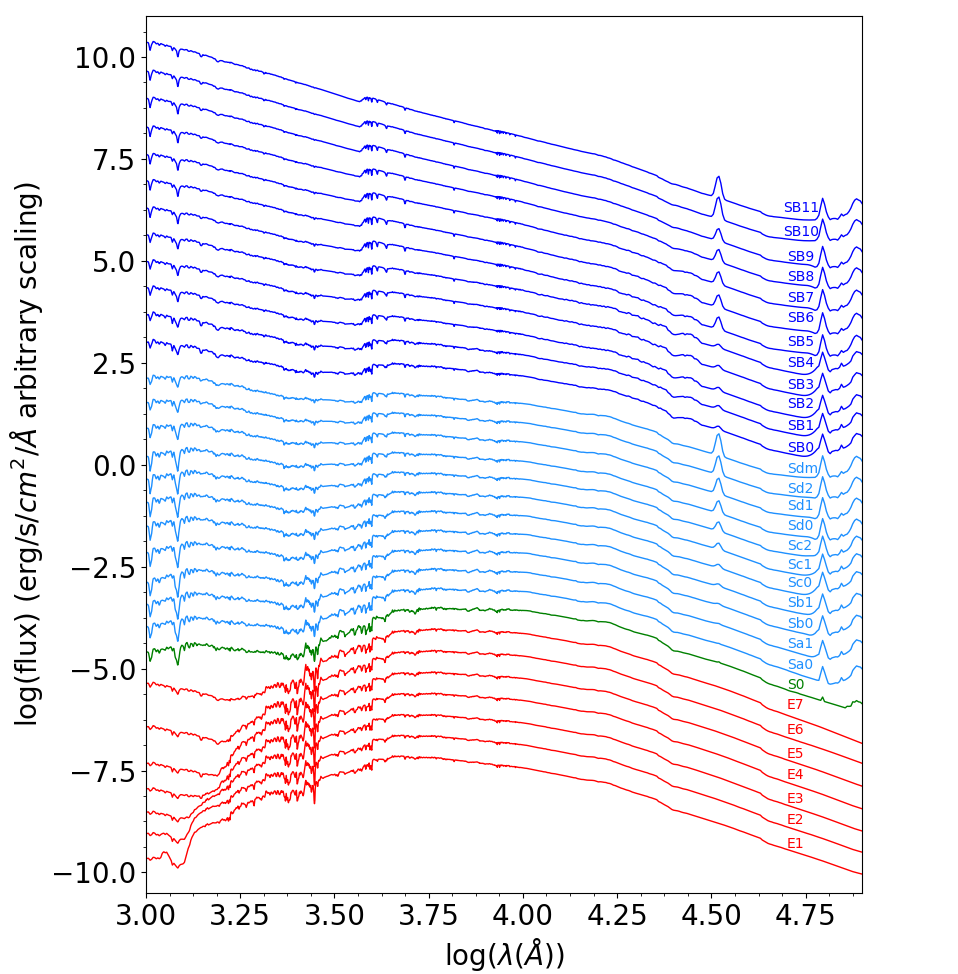}
    
    \caption{SED templates in COSMOS used in this paper. The flux scale is arbitrary. Deep blue curves represent barred spiral galaxies. Light blue curves represent spiral galaxies. The green curve represents a lenticular galaxy. The red curves represent elliptical galaxies.}
    \label{fig:tempspectra}
\end{figure}

After the analysis of both model templates and filter combinations in Section \ref{sec:model} and \ref{sec:filtercomb}, we conclude that we have the following best parameters assuming the best photo-z performance (the lowest catastrophic error and photo-z dispersion).
For filter combinations, we use 26 bands from optical to NIR (< 7 $\mu$m) photometric data. They are Maidanak/SNUCAM $B$, $R$, $I$, Subaru/HSC $g$, $r$, $i$, $z$, $Y$, CFHT/MegaCam $u^{*}$, $g$, $r$, $i$, $z$, CFHT/MegaPrime $u$, KPNO/FLAMINGOS $J$, $H$, CFHT/WIRCam $Y$, $J$, $K_s$, $AKARI$/IRC $N2$, $N3$, $N4$, WISE W1, W2, and $Spitzer$/IRAC 1 and 2.
For galaxy templates, we use COSMOS \citep{Ilbert2009}. Following \citet{Ilbert2009}, we apply 4 different extinction laws to the COSMOS SED templates. They are SMC, star burst, SB calzetti with two different types of bumps \citep[See][for more details of these extinction models]{Bolzonella2000}, with reddening E(B-V) of 0.000, 0.100, 0.200, 0.300, 0.400, 0.500 for each extinction law. \citet{Ilbert2009} found out COSMOS SEDs have the best performance with the above configuration. The parameter set-up of this work is summarised in Table \ref{tab:para}. Statistics of different templates used when running photo-z with the whole catalogue is summarised in Fig.~\ref{fig:templatehist.pdf} and Table~\ref{tab:COSMOS}. Spectra of the templates are presented in Fig.~\ref{fig:tempspectra}.

\begin{figure*}
\centering
	\includegraphics[width=120mm]{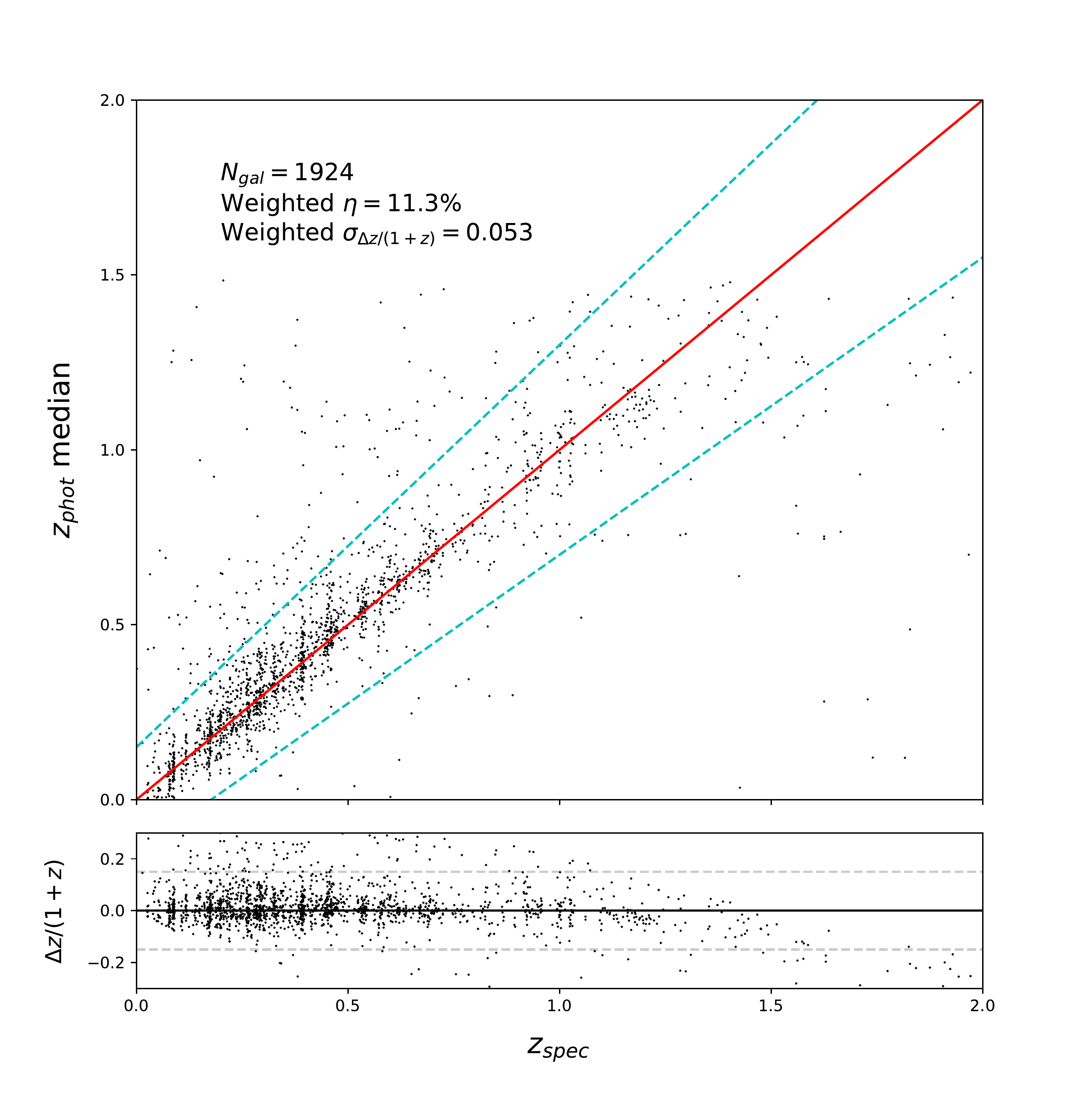}
    
    \caption{Top panel: The photo-z versus spec-z diagram at the NEPW field. Bottom panel: $\Delta z/(1+z)$ versus spec-z diagram at photo-z < 1.5.}
    \label{fig:photoz}
\end{figure*}

\begin{figure*}
\centering
    \includegraphics[width=145mm]{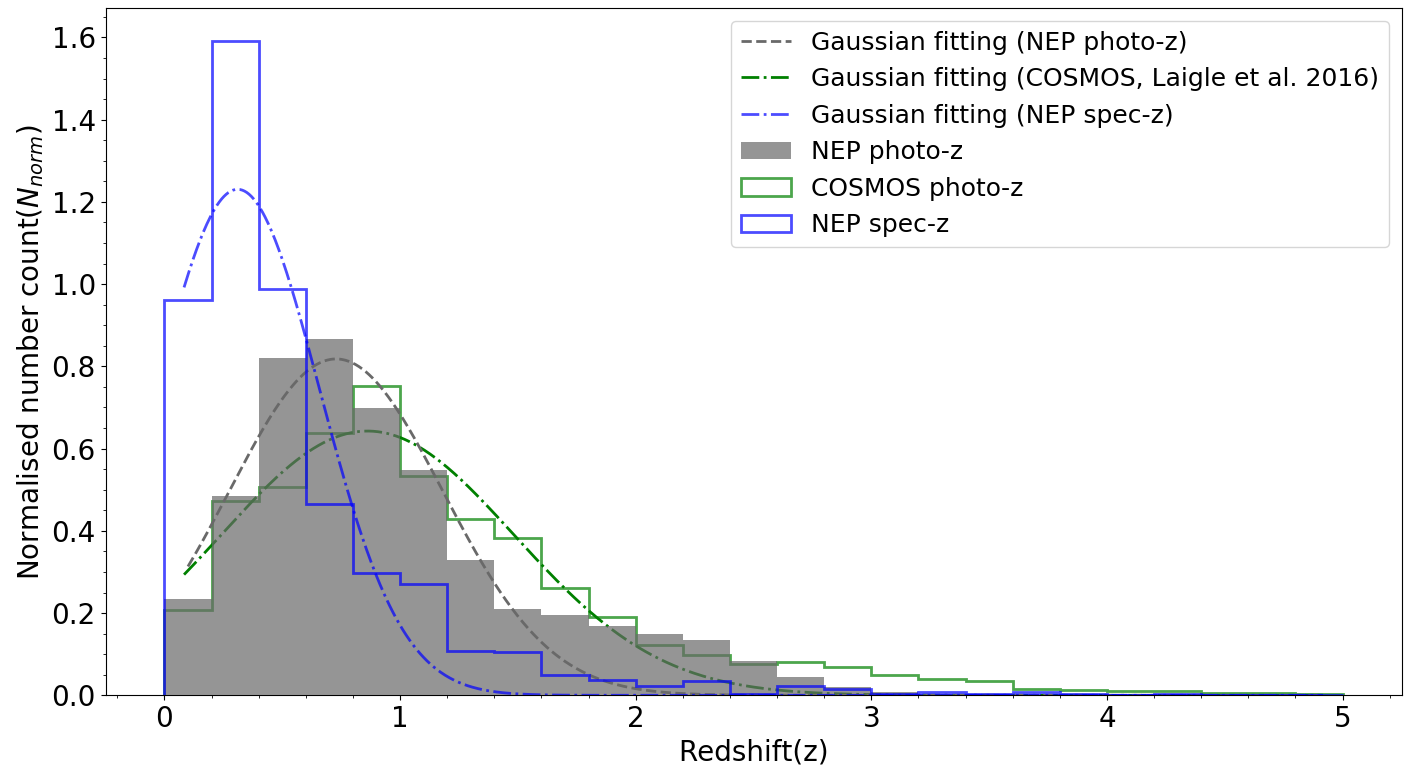}
    
    \caption{Histogram of photo-z distribution in grey. Also shown are Gaussian fits, and spec-z distribution histograms in blue. The COSMOS photo-z histogram is also plotted in green along with its Gaussian fitting. The histograms are normalised to area = 1.}
    \label{fig:photozhist}
\end{figure*}

At photo-z < 1.5, we reach a weighted redshift dispersion of {$\sigma_{\Delta{z/(1+z)}}$} = 0.053. The weighted catastrophic error rate is $\eta$ = 11.3\%. Fig. \ref{fig:photoz} shows the photo-z versus spec-z diagram. Fig. \ref{fig:photozhist} shows the photo-z distribution histogram for the whole catalogue. There are 78,574 sources for z < 1.5. We also plot the redshift distribution of the COSMOS field \citep{Laigle2016}. There are many more sources in COSMOS because they have much deeper data than ours ($\sim$2 mag). We limit the HSC $i$-band magnitude to its depth, 25.97 (mag) and fit the normalised histograms with Gaussian. For NEP photo-z, it has a peak position of (z, $N_{norm}$)=(0.728, 0.818), which $N_{norm}$ is the normalised number count, and a standard deviation ($\sigma$) of 0.454. For COSMOS photo-z, it has a peak position of (z, $N_{norm}$)=(0.873, 0.643) and a $\sigma$ of 0.618. COSMOS's photo-z has a larger redshift peak position and a larger $\sigma$ in its redshift range. To explain this, COSMOS has a smaller field (2.0 deg$^{2}$) coverage compared with NEP (5.4 deg$^{2}$). Therefore, COSMOS might be affected by large scale structures in the field area and thus the redshift range is different from NEP.  

From Table \ref{tab:models}, and Table \ref{tab:filtercomb}, most of our set-ups show a better photo-z performance after weighting. In general, we should have worse photo-z for faint or higher redshift galaxies. However, this is not the case in the NEP with our 26-band photometry. In our paper, the major optical data are the HSC and >95\% of the sources have detections in HSC. We used 5 bands in HSC, which are $g$, $r$, $i$, $z$, and $Y$. They cover wavelengths of $\sim$0.4-1 $\mu$m. Photo-z performance is very good at z$\sim$0.4-1.5 because we can capture the Balmer break using the HSC photometry within this redshift range. Therefore, these fainter galaxies have good photo-z estimation, and improve the accuracy after weighting within z$\sim$0.4-1.5. For brighter galaxies at z$\sim$0-0.4, we need $u$-band data to capture the Balmer break. However, our MegaPrime/MegaCam $u$-band data do not cover the whole NEPW field (<40\% with detections). Even when they do, it is not deep enough compared to the HSC data. Thus, brighter galaxies at z$\sim$0-0.4 sometimes have worse photo-z. We have checked the photo-z dispersion and catastrophic errors in smaller redshift bins i.e. z=0.0-0.2, z=0.2-0.4... The photo-z dispersion and catastrophic error indeed worsen after weighting at z<0.4. On the other hand, photo-z dispersion and catastrophic error improve in z$\sim$0.4-1.5 after weighting. Then, they become worse again at z>1.5 as the Balmer break is redshifted to $J$-band.

We provide the probability distribution function (PDF,  defined as $\int_{}F(z)dz$ between $z_{best}\pm0.1(1+z)$) for all photo-z in the online catalogue (PDF in Table \ref{tab:catalog}, so that users can decide their sample selection criteria based on the PDF.)

\begin{table*}
	\centering
	\caption{The best parameter set-up we used for the computation of photo-z in $Le$ $Phare$.}
	\label{tab:para}
	\begin{tabular}{cc} % four columns, alignment for each
		\hline\hline
		Parameter &  Set-up\\
		\hline
		Galaxy template model & COSMOS\char`_MOD \\
		Z\char`_STEP & 0.01 \\
		Z\char`_MAX & 5.0 \\
		EXTINC\char`_LAW & SMC\char`_prevot.dat, SB\char`_calzetti.dat, SB\char`_calzetti\char`_bump1.dat, SB\char`_calzetti\char`_bump2.dat \\
		E(B-V) & 0.000, 0.100, 0.200, 0.300, 0.400, 0.500 \\
				
		\hline
	\end{tabular}
\end{table*}

\begin{table}
\centering
	\caption{The list of galaxy templates in COSMOS \citep{Ilbert2009}. The morphology is classified based on Hubble-de Vaucouleurs galaxy morphological classification. The spectra are shown in Fig. \ref{fig:tempspectra}.}
	\label{tab:COSMOS}
	\begin{tabular}{ccc} % four columns, alignment for each
		\hline
		No. & Template name & Morphology\\
		\hline
		 1 & Ell1\char`_ A\char`_ 0 & E1 Elliptical galaxy$^{a}$ \\ 
         2 & Ell2\char`_ A\char`_ 0 & E2 Elliptical galaxy$^{a}$ \\   
		 3 & Ell3\char`_ A\char`_ 0 & E3 Elliptical galaxy$^{a}$ \\   
		 4 & Ell4\char`_ A\char`_ 0 & E4 Elliptical galaxy$^{a}$ \\   
		 5 & Ell5\char`_ A\char`_ 0 & E5 Elliptical galaxy$^{a}$ \\   
		 6 & Ell6\char`_ A\char`_ 0 & E6 Elliptical galaxy$^{a}$ \\   
		 7 & Ell7\char`_ A\char`_ 0 & E7 Elliptical galaxy$^{a}$ \\   
		 8 & S0\char`_ A\char`_ 0 & Lenticular galaxy (S0) \\     
		 9 & Sa\char`_ A\char`_ 0 & Sa Spiral galaxy (1)$^{b}$ \\     
		 10 & Sa\char`_ A\char`_ 1 & Sa Spiral galaxy (2)$^{b}$ \\     
		 11 & Sb\char`_ A\char`_ 0 & Sb Spiral galaxy (1)$^{b}$ \\     
		 12 & Sb\char`_ A\char`_ 1 & Sb Spiral galaxy (2)$^{b}$ \\     
		 13 & Sc\char`_ A\char`_ 0 & Sc Spiral galaxy (1)$^{b}$ \\     
		 14 & Sc\char`_ A\char`_ 1 & Sc Spiral galaxy (2)$^{b}$ \\     
		 15 & Sc\char`_ A\char`_ 2 & Sc Spiral galaxy (3)$^{b}$ \\     
		 16 & Sd\char`_ A\char`_ 0 & Sd Spiral galaxy (1)$^{b}$ \\     
		 17 & Sd\char`_ A\char`_ 1 & Sd Spiral galaxy (2)$^{b}$ \\     
		 18 & Sd\char`_ A\char`_ 2 & Sd Spiral galaxy (3)$^{b}$ \\     
		 19 & Sdm\char`_ A\char`_ 0 & Sm Spiral galaxy$^{b}$ \\    
		 20 & SB0\char`_ A\char`_ 0 & Barred Spiral galaxy (1) \\    
		 21 & SB1\char`_ A\char`_ 0 & Barred Spiral galaxy (2) \\    
		 22 & SB2\char`_ A\char`_ 0 & Barred Spiral galaxy (3) \\    
		 23 & SB3\char`_ A\char`_ 0 & Barred Spiral galaxy (4) \\    
		 24 & SB4\char`_ A\char`_ 0 & Barred Spiral galaxy (5) \\    
		 25 & SB5\char`_ A\char`_ 0 & Barred Spiral galaxy (6) \\    
		 26 & SB6\char`_ A\char`_ 0 & Barred Spiral galaxy (7) \\    
		 27 & SB7\char`_ A\char`_ 0 & Barred Spiral galaxy (8) \\    
		 28 & SB8\char`_ A\char`_ 0 & Barred Spiral galaxy (9) \\    
		 29 & SB9\char`_ A\char`_ 0 & Barred Spiral galaxy (10) \\    
		 30 & SB10\char`_ A\char`_ 0 & Barred Spiral galaxy (11) \\   
		 31 & SB11\char`_ A\char`_ 0 & Barred Spiral galaxy (12) \\ 		  
		\hline
	\end{tabular}\\
	\begin{flushleft}
	(a) Elliptical galaxies are labelled En, where the number n is determined by the galaxy shape, where n = 10(1-B/A). (A and B are the semi-minor and semi-major axis respectively.)\\
	(b)Spiral galaxies are classified as Sa, Sb, Sc, and Sm. They have a decreasing bulge-to-disk luminosity respectively.
	\end{flushleft}
\end{table}

\section{Discussion}\label{sec:discuss}

\subsection{Photo-z of AGN}
An AGN is a luminous compact region at the centre of a galaxy. The SED of a galaxy with an AGN differs from that of normal SFGs due to the strong radiation from the accretion disk around the supermassive black hole, and higher dust temperature. Although AGN are very bright, they may be obscured by gas and dust. The obscured AGN could be missed by optical observation. Therefore, MIR data is more important to overcome this problem \citep{Toba2015, Toba2016, Goto2019}.

In this section, we discuss the effect of AGN on photo-z performance. In Section \ref{sec:model}, we consider the effect of AGN in our catalogue. We already tested the addition of Polletta's AGN templates into COSMOS, to see if we achieve a better performance than Polletta's template or the COSMOS template. From Table. \ref{tab:models}, the templates without AGN perform better. This is probably because including AGN templates causes additional misclassification by $Le$ $Phare$, i.e., in some cases, non-AGN would be fitted by AGN templates. As a result, the entire performance of photo-z could be worsen. Due to this reason, we do not consider the AGN template in our photo-z estimation. At this moment, we do not have deep enough X-ray data. In the future, one way to improve is taking a two-step approach of separating AGN from galaxies by using X-ray identified AGN. Then, we perform the SED fitting separately for the AGN and galaxies using different templates. For sources that are not detected in X-rays, we fit them with the galaxy templates. For AGN, we fit galaxy+AGN templates to X-ray sources.

In a different approach, we remove known AGN from our spec-z data to check the performance of photo-zs. In our band-merged photometric catalogue (Kim et al. 2020, in press), spec-z sources have already been classified by the X-ray catalogue from Chandra NEP deep survey \citep{Krumpe2015}. There are flags indicating whether these sources are AGN. In addition, \citet{Shim2013} spectroscopically identified 198 type-1 AGN, and 8 type-2 AGN in the spectroscopic sample. Therefore, we remove these AGN from our input catalogue and compute photo-z again. In this test, we reach a weighted {$\sigma_{\Delta{z/(1+z)}}$} = 0.053, and weighted $\eta$ = 9.4\% with 1723 sources at z < 1.5.

Compared with the photo-z accuracy computed by $Le$ $Phare$ for the full sample, the weighted {$\sigma_{\Delta{z/(1+z)}}$} does not change, and the catastrophic error only slightly improves by $\sim$2\%. However, we also lose 198 sources' photo-z, which is more than 10\% of sources from the spec-z data. AGN sources do negatively affect photo-z performance. This is understandable as we do not use any AGN template. However, we have decided to keep the AGN in our sample as we want to keep the completeness of our data.

Due to this reason, we do not consider the AGN template in our photo-z estimation. At this moment, we do not have deep enough X-ray data. In the future, one way to improve is taking a two-step approach of separating AGN from galaxies by using X-ray identified AGN. Then, we perform the SED fitting separately for the AGN and galaxies using different templates. For sources that are not detected in X-rays, we fit them with the galaxy templates. For AGN, we fit galaxy+AGN templates to X-ray sources.

\subsection{Galaxy classification in the NEPW field}
Galaxy classification has been one of the most challenging topics in extragalactic astronomy. In this section, we discuss galaxy classification in NEPW field based on $Le$ $Phare$ SED fitting with the COSMOS galaxy SED template.
In Table \ref{tab:COSMOS}, and Fig. \ref{fig:tempspectra}, we present information about different galaxy templates in COSMOS. In Fig. \ref{fig:templatehist.pdf}, the histogram shows the distribution of different galaxy types in the NEPW field. We find $\sim$ 21\%, 5\%, and 74\% for early-types (Template 1-5 in Table \ref{tab:COSMOS}), lenticular (Template 8), star-forming galaxies (Templates 9-31), respectively in our data. Note that the names of the templates in Table \ref{tab:COSMOS} imply morphological classification, but our classification is based on only the SED fitting and should be treated as such.

A large fraction of galaxies in our sample are classified as SFGs. However, this is expected because the SFGs are also luminous in infrared. The domination of SFGs in the sample reflects the characteristics of our $AKARI$ NIR- and MIR-selected galaxies in the NEP field.
Further discussion about galaxy/AGN classification and morphology in NEPW are presented in \citet{Wang2020} and Kim et al. 2020 (in press), respectively.

\subsection{Comparison of photo-z accuracy with other work}
\begin{figure*}
\centering
    \includegraphics[width=100mm]{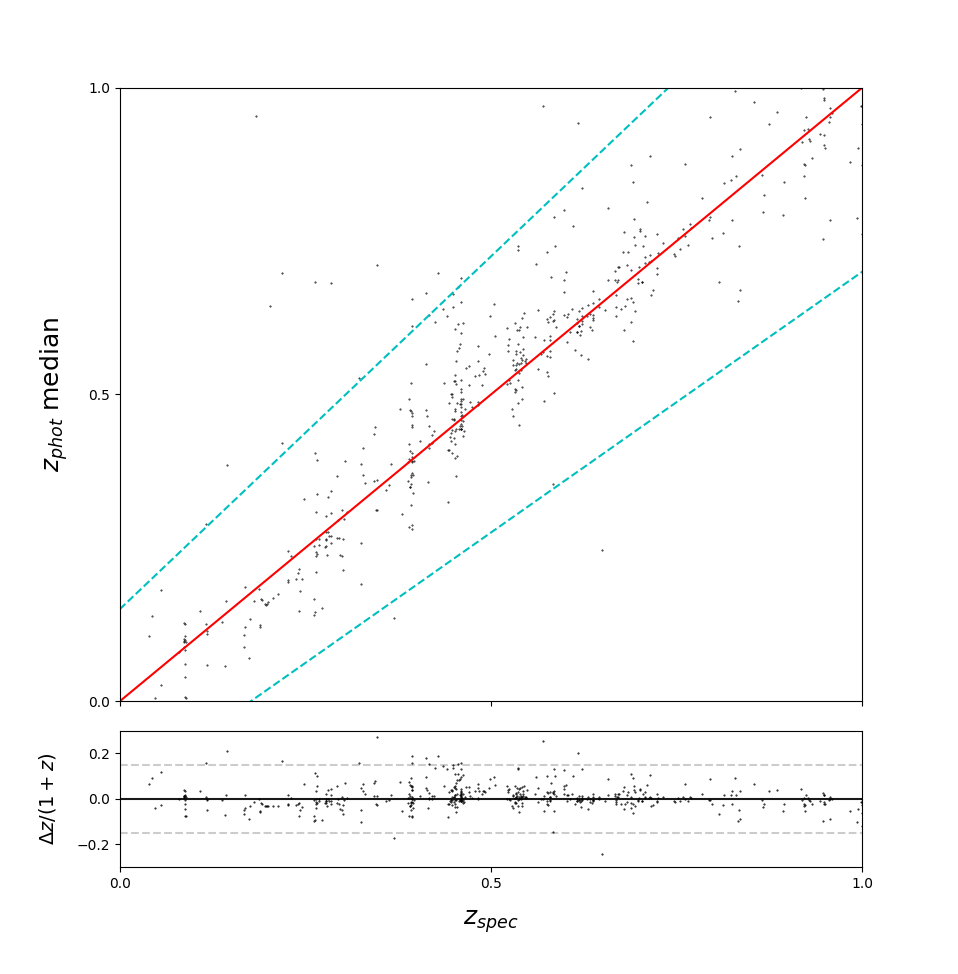}
    
    \caption{Top panel: The photo-z versus spec-z diagram at the NEPD field. Bottom panel:$\Delta z/(1+z)$versus spec-z diagram at both z<1.0}
    \label{fig:deepphotoz}
\end{figure*}

In this section, we compare the photo-z accuracy with other works in the NEP field (NEPD/NEPW). \citet{Oi2014} used 8-band photometry and reach a photo-z dispersion of 0.032 and a catastrophic failure rate of 5.8\% at z < 1. Compared with our work, which reaches a photo-z dispersion of 0.046 [0.047] and a catastrophic failure rate of 9.0\% [9.0\%] for z < 1, their results performed slightly better. However, the data they used were mostly from NEPD survey. The NEPW survey covers 5.4 deg$^2$, but the NEPD survey covers only 0.67 deg$^2$. Nevertheless, all NEPD data are deeper because some telescopes only cover the NEPD region e.g. CFHT/WIRCAM.On the other hand, NEPW field is a wider area ($\sim$6,000 more sources) with the combination of swallow and deep data. In terms of spec-z data, \citet{Oi2014} used only 483 spec-z sources to assess the photo-z dispersion at z < 1. In our work, we assess photo-z dispersion with 1,694 for z < 1, which is $\sim$ 3 times more than the previous work. For a fair comparison, we have performed further analysis by limiting our samples to the NEPD field. Our photo-z performs almost the same at z<1.0 in photo-z dispersion (Oi 2014: 0.032, Our work: 0.036) but improves with more than 1\% in the catastrophic outlier fraction significantly. This is mainly due to the fact that we have included deeper Subaru/HSC optical data, while \citet{Oi2014} only used CFHT/MegaCam optical data (i.e., no Subaru/HSC data).
In addition, \citet{Oi2014} did not account for the effect of spectroscopic catalogue bias in their work. Therefore, we believe our weighted result is more representative of the actual data. We summarise the comparisons in Table ~\ref{tab:comparwork}. The photo-z versus spec-z diagram of this work at the NEPD field for z < 1.0 is shown in Fig. \ref{fig:deepphotoz}. Note that the redshift cut we applied in this comparison is both spec-z and photo-z < 1.0, which is the same as \cite{Oi2014}.

%Moreover, Oi et al. in prep. also computed photo-z using NEPW data. They reached a photo-z dispersion of 0.060 and catastrophic failure rate of 8.3\% for z < 1.5 after 10 times 3-$\sigma$ clipping. In this case, our work, which reaches a photo-z dispersion of 0.0493 and catastrophic failure rate of 9.27\% for z < 1.5, performs better in photo-z dispersion. Also, the total number of sources in our work (91,861) is also more than Oi et al. in prep. (89,178). We summarise the comparisons in Table ~\ref{tab:comparwork}.

\begin{table*}
	\centering
	\caption{Summary of the best photo-z performance in this work. Different photo-z work on the deep field is also included. N is the fraction of objects with spec-z over the total number of objects. The weighted photo-z performance is shown in square brackets.}
	\label{tab:comparwork}
	\resizebox{\textwidth}{!}{
	\begin{tabular}{lllllll} % four columns, alignment for each
        \hline
        Work & Redshift & N & $\eta$ & {$\sigma_{\Delta{z/(1+z)}}$} & Template & (No. of filters) Filter combination\\
        \hline
        This work & 0<z<1 & 1683/ & 9.0\% & 0.046 & COSMOS \citep{Ilbert2009} & (26) Maidanak/SNUCAM $B$, $R$, $I$,\\ (NEPW) && 91,861&[9.0\%] & [0.047]
        && Subaru/HSC $g$, $r$, $i$, $z$, $Y$,\\
        &&&&&& CFHT/MegaCam $u^{*}$, $g$, $r$, $i$, $z$,\\
        &&&&&& CFHT/MegaPrime $u$, KPNO/FLAMINGOS $J$, $H$,\\
        &&&&&& CFHT/WIRCam $Y$, $J$, $K_s$,\\
        &&&&&& $AKARI$ $N2$, $N3$, $N4$,\\
        &&&&&& $Spitzer$/IRAC 1, 2, WISE 1, 2\\
        \hline
        This work & 0<z<1 & 517/ & 4.4\% & 0.036 & COSMOS \citep{Ilbert2009} & (26) Maidanak/SNUCAM $B$, $R$, $I$,\\ (NEPD) && 91,861&[4.6\%] & [0.045]
        && Subaru/HSC $g$, $r$, $i$, $z$, $Y$,\\
        &&&&&& CFHT/MegaCam $u^{*}$, $g$, $r$, $i$, $z$,\\
        &&&&&& CFHT/MegaPrime $u$, KPNO/FLAMINGOS $J$, $H$,\\
        &&&&&& CFHT/WIRCam $Y$, $J$, $K_s$,\\
        &&&&&& $AKARI$ $N2$, $N3$, $N4$,\\
        &&&&&& $Spitzer$/IRAC 1, 2, WISE 1, 2\\  
        \hline
        \citet{Oi2014} & 0<z<1 & 483/ & 5.8\% & 0.032 & \citet{Coleman1980} & (8) CFHT/MegaCam $u^{*}$, $g'$, $r'$, $i'$, $z'$,\\ (NEPD) && 85,797 &&& \citet{Kinney1996} & CFHT/WIRCam $Y$, $J$, $K_s$,\\
        \hline
	\end{tabular}}
\end{table*}

\section{Conclusions}\label{sec:conclusion}
This paper presents a photo-z catalogue for the IR sources observed by $AKARI$ space telescope and identified by a deep HSC survey on the NEP-Wide survey field. This photo-z catalogue is computed with $Le$ $Phare$ and a total of 26 filters from the optical to NIR (< 7 $\mu$m) photometric data, which are Maidanak/SNUCAM $B$, $R$, $I$, Subaru/HSC $g$,  $r$, $i$, $z$, $Y$, CFHT/MegaCam $u^{*}$, $g$, $r$, $i$, $z$, CFHT/MegaPrime $u$, KPNO/FLAMINGOS $J$, $H$, CFHT/WIRCam $Y$, $J$, $K_s$, $AKARI$/IRC $N2$, $N$3, $N4$, WISE W1, W2, and $Spitzer$/IRAC 1 and 2.
We separate star-like objects from the galaxies by using the $\chi^2_{gal}$ and $\chi^2_{star}$ which are computed from $Le$ $Phare$.There are 24,110 out of 91,861 objects with $\chi^2_{gal}$ > $\chi^2_{star}$. They are flagged as possible stars in this NEP photo-z catalogue (Flag\char`_ 1; Table \ref{tab:catalog}). 
For galaxy templates, we use COSMOS \citep{Ilbert2009} with reddening $E(B-V)$ of 0.000, 0.100, 0.200, 0.300, 0.400, and 0.500. We computed weights by the NN algorithm \citep{Lima2008}. The weight is applied to every spec-z sources in order to match the magnitude distribution of the spec-z and the photometric catalogue. In general, the weighted dispersion and outlier fraction only slightly change at our target z < 1.5 redshift range. Most of them improve their photo-z (photo-z dispersion $\sim$0.003, outlier fraction $\sim$2\%) after applying the weights. For more detail, see Table \ref{tab:models}, and \ref{tab:filtercomb}.
Among all the templates that we tried, COSMOS provides the best performance based on both weighted photo-z dispersion (0.053) and catastrophic error (11.3\%) in 26-band filter combination at z < 1.5.
We use the $\chi^2$ value to separate stars and galaxies. Note that we rely on the SED when separating stars and galaxies. There are 24,110 out of 91,861 objects with $\chi^2_{gal}$ > $\chi^2_{star}$. They are flagged as possible stars in this NEP photo-z catalogue (Flag\char`_ 1; Table \ref{tab:catalog}).
We present and release the photo-zs for 91,861 objects in the NEPW field survey, which covers a total effective area of 5.4 deg$^2$ in redshift range 0 < z < 5. We use 2026 spec-z at 0 < z < 5 from the NEP as a calibration to derive these photo-z. For photo-z < 1.5, we reach a weighted redshift dispersion of {$\sigma_{\Delta{z/(1+z)}}$} = 0.053 with weighted $\eta$ = 11.3\% of catastrophic errors ($\eta$ is defined strictly as those objects with |$\Delta$z|/(1 + $z_s$) > 0.15).

We also conclude that the effect of AGN on photo-z is not significant in our case. Excluding AGN only improves the photo-z accuracy by $\sim$ 0.008 and removes $\sim$ 5 \% of the outliers. Losing 10\% (AGN) of the data is the prerequisite of this photo-z accuracy. Therefore, we keep AGN in our catalogue, so we have more photo-zs and these photo-zs can be used for AGN related research.
%xxx cut. The photo-z for sources with larger magnitude are less reliable compared those with smaller magnitude. However, we will lose nearly half of the sources if we exclude sources with magnitude > 20. Therefore, we did not adopt a magnitude cut. This improve the photo-z performance, and save more sources. 

The full version of these photo-zs 
(Table \ref{tab:catalog}) will be shared on reasonable request to the corresponding author.

\section*{Acknowledgements}
We are very grateful to the anonymous referee for many insightful comments. This research is based on observations with $AKARI$, a JAXA project with the participation of ESA. TG acknowledges the supports by the Ministry of Science and Technology of Taiwan through grants 105-2112-M-007-003-MY3 and 108-2628-M-007-004-MY3. TH is supported by the Centre for Informatics and Computation in Astronomy (CICA) at National Tsing Hua University (NTHU) through a grant from the Ministry of Education of the Republic of China (Taiwan). This research was conducted under the agreement on scientific cooperation between the Polish Academy of Sciences and the Ministry of Science and Technology in Taipei. AP has been supported by the Polish National Science Centre grant UMO-2018/30/M/ST9/00757. This work is partly based on tools and data products produced by GAZPAR operated by CeSAM-LAM and IAP. We greatly appreciate Dr. Olivier Ilbert for many insightful comments and advice on $Le$ $Phare$. We also appreciate Prof. Myungshin Im for providing information on Maidanak/SNUCAM filters and Dr. Dick Joyce for providing information on KPNO/FLAMINGOS filters.

\section*{Data availability}
The data underlying this article will be shared on reasonable request to the corresponding author.
%%%%%%%%%%%%%%%%%%%%%%%%%%%%%%%%%%%%%%%%%%%%%%%%%%



\bibliographystyle{mnras}
\bibliography{Simonphotoz}

%\bibliographystyle{mnras}


%%%%%%%%%%%%%%%%%%%%%%%%%%%%%%%%%%%%%%%%%%%%%%%%%%

%%%%%%%%%%%%%%%%% APPENDICES %%%%%%%%%%%%%%%%%%%%%
\appendix
\section{Summary of source detection and photometry from different sources}\label{detection}
To detect sources and conduct photometry, most of the catalogues we used applied the same method \citep[SExtractor;][]{Bertin1996} but they are slightly different in the parameter set up. Note that the detections were performed using the same aperture across the HSC's $g$, $r$, $i$, $z$, and $Y$-bands, but not including other filters in this paper. We present a summary of their source detection and photometry in the following paragraphs.

For Maidanak/SNUCAM, \citet{Jeon2010} stacked $B$-, $R$-, and $I$-band images together, and used the BRI stacked image as their object detection image. They chose a set of 1.2$\sigma$ for DETECT\char`_ THRESH, 5 connected pixels for DETECT\char`_ MINAREA, DEBLEND\char`_ NTHRESH of 64, DEBLEND\char`_ MINCONT of 0.005, 200 for BACK\char`_ SIZE, and 3 for BACK\char`_ FILTERSIZE as the optimal set of detection parameters. The adopted detection limit corresponds to the S/N of about 4.5. They derived aperture magnitudes using an aperture of diameter of three times FWHM and auto-magnitudes with Kron-like elliptical apertures which are taken as the total magnitudes.

For Subaru/HSC, Oi et al. 2020 (in press) used SExtractor for source detection and photometric measurement. However, they used C\char`_ MODEL flux instead of APERTURE\char`_ flux. In \citet{Kim2012}'s $AKARI$ source catalogue, it contains both galaxies (mainly detected in MIR bands) and stars (mainly detected in NIR bands).
The seeing condition for our HSC r-band data is worse compared to the other four bands. To improve the photometry before source matching of $AKARI$ and HSC data, they decided to use the C\char`_ MODEL photometry.

For MegaCam and WIRCam, \citet{Oi2014} carry out the photometry for each band with optimal aperture. Then, they ran SExtractor with each band image individually. Fluxes of extracted sources were measured using SExtractor with elliptical apertures of 2.5 times of the Kron radius.

For CFHT/MegaPrime, \citet{Huang2020} used SExtractor to extract sources from an image and performs photometry of the sources. They chose to use a filter with a narrow Gaussian FWHM and a small convolution mask in order not to miss any possible faint sources. They had compared the source extraction results using several filters, and eventually they adopted the result using Gaussian PSF with 2-pixel FWHM and the 3-pixel by 3-pixel convolution mask (Gauss2.0 3x3).

For $AKARI$/IRC, \citet{Kim2012} used the entire mosaicked images covering the whole NEPW area to carry out extraction and photometry of the sources for each band. To measure the fluxes of the detected sources, they used a SExtractor \citep{Bertin1996} and ran it in single mode. They chose DETECT\char`_ THRESH = 3, DETECT\char`_ MINAREA = 5, and BACK\char`_ SIZE = 3, which are the same as those used by \citet{Wada2008}, except for DETECT\char`_ THRESH. They chose a higher threshold to reduce false detections.

For KPNO/FLAMINGOS, \citet{Jeon2014} chose the following set of detection parameters for SExtractor: DETECT\char`_ MINAREA = 20, DETECT\char`_ THRESH = 1.1, DEBLEND\char`_ NTHRESH = 64, DEBLEND\char`_ MINCONT = 0.005, BACK\char`_ SIZE = 500, BACK\char`_ FILTERSIZE = 2. The adopted detection parameters correspond to S/N = 5. They derived auto magnitudes with Kron-like elliptical aperture, which are assumed to be total magnitudes. For the aperture magnitudes, they set the aperture diameter size to three times the FWHM.

For $Spitzer$/IRAC, \citet{Nayyeri2018} ran SExtractor in dual mode on the combined three-epoch mosaics on the 3.6 $\mu$m and 4.5 $\mu$m individually, as the detection bands with photometry are extracted in both. The two catalogues are then merged to form the final NEP catalogue. They measured source photometry in AB magnitudes given the pixel flux units in micro-Jansky and reported this in a Kron radius (the MAG\char`_ AUTO), which is photometry over an ellipse with its size and orientation determined from the second moment of light distribution above the isophotal threshold, and also over two apertures (with 4" and 6" diameter).

For WISE, \citet{Jarrett2011} did not use SExtractor but carried out extraction on each individual frame for all bands with their own algorithm simultaneously. It utilises an optimal combination of the multi-band imaging data for source photometry, mitigating unruly pixels, artefacts, cosmic radiation hits, source saturation, and angular resolution differences. They did not mention how source magnitude are measured.

\section{Summary of definitions of depths for different catalogues}\label{depth}
Most of the bands we used throughout this paper have a similar definition when defining 5$\sigma$ limiting magnitude. The others used apertures of different sizes to measure the limiting magnitudes from the sky background. We present a summary of their methodology in the following paragraphs.

For Maidanak/SNUCAM \citep{Jeon2010} and KPNO/FLAMINGOS \citep{Jeon2014}, the 5$\sigma$ detection limits of each subfield were determined from the sky background $\sigma$, within an aperture diameter of three times the FWHM value. The detection limits vary from subfield to subfield. They measured 5$\sigma$ detection limits for each subfield. In their paper, they did not mention the size of the aperture. In this paper, we present the median value of the 5$\sigma$ detection limits for all subfields from \citet{Jeon2010} and \cite{Jeon2014}.

For Subaru/HSC (Oi et al. 2020, in press), they determined the 5$\sigma$ detection limits over a circular aperture of 6.0 pixels radius which corresponds to 1.02 arcsec. In addition, they calculated the S/N ratio of all the sources in a patch from their fluxes and flux errors within the 6.0 pixel aperture. They performed this process for all the 251 patches. Since the number of coadded images differ depending on the sky region due to the HSC FoV positions in the observation, the 5$\sigma$ limiting magnitudes vary depending on patches. In this paper, we present the median values of the  5$\sigma$ limiting magnitudes from all patches.

For CFHT/MegaCam \citep{Oi2014} and CFHT/WIRCam \citep{Oi2014}, They determined the detection limits over a circular aperture of 1 arcsec radius. The detection limit changes depending on the source position on the mosaiced images since the exposure map is not uniform. We present the median value in Table \ref{tab:filterinfo}.

For CFHT/MegaPrime $u$ \citep{Huang2020}, they estimated the depth by placing 20,000 2 arcsec apertures randomly in the co-added images. Since the pixel value of the co-added image is calibrated with the zero point, they converted the 5$\sigma$ depth by using the equation, MAG (depth) = 25.188 - 2.5 x log(5$\sigma$f), where the $\sigma$f is the standard deviation obtained from the Gaussian fitting of the distribution of the 20,000 random aperture flux values.

For $AKARI$/IRC \citep{Kim2012}, the flux limit of the point source detection in each band was estimated from the fluctuation in the sky background, by measuring the flux at random positions far away from the source positions. We used an aperture three times the size of the FWHM, and determined the 5$\sigma$ detection limits based on the value of $\sigma$ derived from the sky background. The detection limits depend on the noise levels of the fields, which vary from place to place, and we present the averaged values over the entire NEP field.

For $Spitzer$/IRAC \citep{Nayyeri2018}, they measured limiting magnitudes over an aperture of 4 arcsec with aperture corrections applied, see detail in \citet{Nayyeri2018}. They corrected the aperture photometry measurements in the 3.6 and 4.5 $\mu$m bands for the effect of fixed apertures (at 4 arcsec and 6 arcsec) used in their measurements.

For WISE \citep{Jarrett2011}, they used a different method from those mentioned above to determine limiting magnitudes. The depths of their combined mosaics was much deeper than the typical WISE field, which has a depth of coverage from $\sim$ 10-15 passes, depending on the ecliptic latitude of the field. For both the EP-N and EP-S WISE surveys, they determined the 5$\sigma$ limiting magnitudes by using a photometric error model in SExtractor \citep{Bertin1996}.

%%%%%%%%%%%%%%%%%%%%%%%%%%%%%%%%%%%%%%%%%%%%%%%%%%
% Don't change these lines
\bsp	% typesetting comment
\label{lastpage}
\end{document}